\documentclass[journal=jpcafh,manuscript=article]{achemso}
\usepackage{longtable}
\usepackage{multirow}
\usepackage{amsmath}
\usepackage[usenames,dvipsnames]{xcolor}
\usepackage{soul}
\usepackage{bm}

\usepackage{xr}
\usepackage{amssymb}
\usepackage{siunitx}
\usepackage{threeparttable}
\usepackage{multicol} \usepackage{wrapfig} \usepackage{enumitem}
\usepackage{bm} \usepackage{outlines} 
\usepackage{siunitx}\usepackage{booktabs}
\usepackage{caption}
\usepackage{subcaption}
\usepackage{subfig}
\usepackage{longtable}
\usepackage{multirow}
\usepackage{booktabs}
\usepackage{amsmath}

\usepackage[normalem]{ulem} 

\usepackage{nicematrix}
 
\SectionNumbersOn

\author{Silvan K\"aser} \affiliation[University of Basel]{Department
  of Chemistry, University of Basel, Klingelbergstrasse 80, CH-4056
  Basel, Switzerland.}
  
\author{Jeremy O. Richardson} \affiliation[ETH Zurich]{Department of
  Chemistry and Applied Biosciences, ETH Zurich, 8093 Zurich,
  Switzerland.}
 
\author{Markus Meuwly} \affiliation[University of Basel]{Department of
  Chemistry, University of Basel, Klingelbergstrasse 80, CH-4056
  Basel, Switzerland.}  \email{m.meuwly@unibas.ch}

\title{Accurate Tunneling Splittings for Ever-Larger Molecules from
  Transfer-Learned, CCSD(T) Quality Energy Functions}

\newcommand{\keywordsanie}{
	Computational chemistry \textbullet\ 
	Gas-phase reactions \textbullet\
 	Machine learning \textbullet\ 
	Quantum chemistry \textbullet\ 
	Instanton theory \textbullet\
    Tunneling splittings
}

\makeatother
\begin{document}
\date{\today}

\begin{abstract}
This work combines state-of-the-art machine learning techniques with
highest-level electronic structure calculations and full-dimensional
quantum tunneling calculations to obtain a quantitative
characterization of tunneling splittings for system sizes that are
currently out of reach using traditional approaches. For
intramolecular hydrogen transfer in tropolone, the best computed
splitting including perturbative corrections in the ring-polymer
instanton calculations is 0.94~cm$^{-1}$ and compares with
0.974~cm$^{-1}$ from experiments. On the other hand, for
intermolecular double hydrogen transfer in the (propiolic
acid)--(formic acid) dimer, the computations yield 0.0147 cm$^{-1}$
which is larger by 40 \% compared with experiment (0.0097 cm$^{-1}$)
but still in much better agreement than previous attempts (0.63
cm$^{-1}$). The strategy pursued in the present work is applicable to
yet larger systems and other properties of interest and provides a
rational route for highest-accuracy energy functions for prediction
and benchmarking electronic structure methods vis-a-vis experiments.
\end{abstract}

\section{Introduction}
Quantum mechanical tunneling is a central concept in chemistry and
physics.\cite{gol:2021} One of the experimentally observable hallmarks
of tunneling processes is the splitting of energy levels. These
tunneling splittings are measurable and can be probed, \textit{e.g.},
using microwave\cite{rowe1976intramolecular} or infrared
spectroscopy.\cite{ortlieb2007proton} The splittings are exquisitely
sensitive to the energetics and shape of the underlying potential
energy surface (PES) and provide valuable benchmarks to validate
state-of-the-art theoretical methodology.\\

\noindent
Accurate PESs allow to carry out quantitative molecular simulations
that link directly to experiments. One of the most successful recent
approaches to devise such PESs uses a combination of electronic
structure calculations and machine learning techniques to give a
machine learning-based PES. Coupled cluster-based techniques such as
CCSD(T) are among the most accurate methods for closed-shell
systems. However, due to the unfavourable scaling of such calculations
(seventh power of the number of basis functions), the necessary number
of reference calculations required for training a full-dimensional PES
($\sim 10^4$ or more) at this level of theory is only feasible for
small molecules with $\lesssim 5$ heavy atoms. However, if a small
number of calculations can be afforded at such high levels, a
considerable boost in performance can be achieved using transfer
learning (TL)
techniques.\cite{taylor2009transfer,smith2019approaching,pan2009survey}
The essence of TL is that the overall topographies of molecular PESs
at different levels of theory are often comparable which can be
exploited by retraining a low-level (LL) PES with small amounts of
data computed at a high level. Such an approach is explored in the
present work for two systems that have so far been intractable at the
CCSD(T) level of theory.\\

\noindent
Computing accurate tunneling splittings for multidimensional systems
from fully quantum-mechanical methods is a formidable task. The
ring-polymer instanton (RPI) approach provides a semiclassical
approximation for tunneling calculations which scales well with system
size.\cite{tunnel,InstReview,hexamerprism} As instanton theory makes a
local harmonic approximation around the optimal tunneling pathway, it
typically deviates from quantum-mechanical benchmarks by up to 
20\%.\cite{Perspective} This error can, however, be significantly
reduced using a new procedure to perturbatively correct the results
using information about third and fourth derivatives along the
path.\cite{richardson2023pcrpi}\\

\begin{figure}[t!]
\centering
\includegraphics[width=0.9\textwidth]{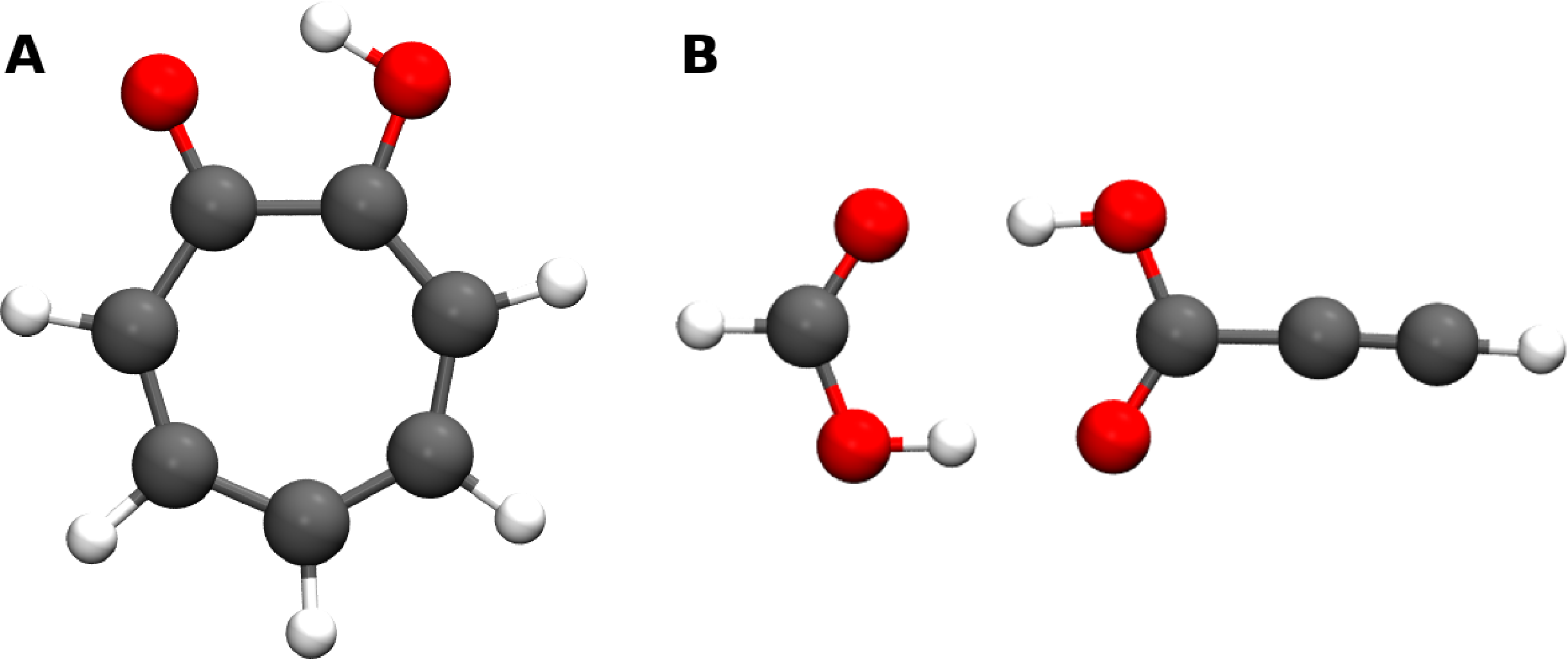}
\caption{Global minimum energy structures for tropolone (A) and PFD (B).}
\label{fig:molecules}
\end{figure}

\noindent
The present work reports tunneling splittings for 15-atom tropolone
and 12-atom (propiolic acid)--(formic acid) dimer (PFD) (see
Figure~\ref{fig:molecules}) following the TL+RPI
approach.\cite{mm.tlrpimalonaldehyde:2022} These molecules are related
to malonaldehyde
\cite{firth1991tunable,baba1999detection,baughcum1984microwave,
  turner1984microwave,smith1983infrared,firth1989matrix,
  chiavassa1992experimental,duan2004high,wang2008full,MM.ma:2010,
  mizukami2014compact,MM.ma:2014,cvitas2016locating,mm.ht:2020,jahr2020instanton,mm.tlrpimalonaldehyde:2022,richardson2023pcrpi}
and the formic acid dimer
(FAD)\cite{ito2000jet,georges2004jet,zielke:2007,ortlieb2007proton,
  xue2009probing,suhm:2012,goroya2014high,duan:2017,li2019barrier,suhm:2020,
  kalescky2013local,ivanov2015quantum,miliordos2015validity,tew2016ab,qu2016ab,MM.fad:2016,richardson2017full,qu2018high,qu2018quantum,qu2018ir},
which have been studied experimentally and theoretically in great
detail. Both molecules have been treated with RPI, which was
particularly successful for malonaldehyde. Using perturbatively
corrected RPI on two different malonaldehyde
PESs\cite{richardson2023pcrpi,kaeser2023numerical} gave a tunneling
splitting for hydrogen (H-)transfer of 22.1~cm$^{-1}$ whereas the
experimental value is 21.6 cm$^{-1}$. Compared with malonaldehyde and
FAD, far less theoretical and experimental work has been devoted to
their ``larger siblings'', tropolone and PFD, mostly due to the
difficulty of obtaining CCSD(T) quality PESs for these systems of many
electrons.\\

\noindent
The experimentally reported ground vibrational state (H/D) tunneling
splittings for tropolone are
(0.974/0.051)~cm$^{-1}$.\cite{tanaka1999determination,keske2006highresolution}
Theoretical works include a permutationally invariant polynomial PES
based on 6604 B3LYP reference energies and
forces.\cite{houston2020permutationally} This was recently improved to
near-CCSD(T) quality using the molecular tailoring approach
(MTA)\cite{sahu2014molecular} to obtain approximate CCSD(T) energies
and $\Delta$-machine learning based on 2044
energies.\cite{nandi2023ring} The (H/D) tunneling splitting as
evaluated on the $\Delta$-machine learned PES was
(0.68/0.031)~cm$^{-1}$, which was subsequently improved by scaling the
action using further MTA calculations to give
(0.92/0.042)~cm$^{-1}$.\\

\noindent
For PFD, the experimental energy splittings are
(0.009721/0.0001117)~cm$^{-1}$ for H-H and D-D exchange,
respectively.\cite{daly2011microwave} Theoretically, PFD has, for
example, been investigated using second-order vibrational perturbation
theory\cite{meyer2021cc} and a one-dimensional model has been employed
to estimate tunneling splittings\cite{sun2013calculations}, which,
however, were not in quantitative agreement with experiment. For
instance, the (H/D) tunneling splittings based on a rectilinear path
determined from a MP2/aug-cc-pVDZ intrinsic reaction coordinate scan
were (0.63/0.0245)~cm$^{-1}$, and thus required a dramatic empirical
scaling in a subsequent step.\\

\section{Results and Discussion}
For both tropolone and PFD, a sufficiently large number of
CCSD(T)/aug-cc-pVTZ calculations is currently out of reach
(\textit{e.g.}, for tropolone: ``Ideally, we would learn the
difference between the DFT and the “gold standard” CCSD(T)
level. However, this is not feasible (for us) owing to $N^7$ scaling
...''\cite{nandi2023ring}). In an attempt to reach CCSD(T) quality,
the present work employs TL strategies for tropolone and PFD. First,
LL reactive PESs were determined, see Supporting
Information. Schematics of the two PESs at the MP2 (low-)level
including molecular structures and the accuracy of the energy on
stationary points are shown in Figures~\ref{sifig:pes_tropo_ll} and
\ref{sifig:pes_PFD_ll}. Comparing the energies of the stationary
points of the PESs with direct \textit{ab initio} calculations yield
maximum deviations of 0.05 and 0.03~kcal/mol for tropolone and PFD,
respectively. Additional validations on a hold-out (test) set, see
Table~\ref{tab:ll} and Figure~\ref{fig:ll_corr}, feature mean absolute
errors on energies (MAE($E$)) well below 0.1~kcal/mol.\\

\begin{table}[h]
\begin{tabular}{crr}\toprule
kcal/mol(/\AA) &  \textbf{Tropolone}   &  \textbf{PFD}\\\midrule
MAE($E$)      &  0.067 & 0.034  \\
RMSE($E$)     &  0.094 & 0.165 \\
MAE($F$)      &  0.069 & 0.133 \\
RMSE($F$)     &  0.336 & 1.793 \\\midrule
$E_{\rm B}$   & 4.284  & 6.460 \\
$E_{\rm B}^{\rm Ref.}$  & 4.285 & 6.470 \\\midrule
MAE($\omega$)$^{\rm min}$ (cm$^{-1}$)      &  1.02 & 0.37  \\
MAE($\omega$)$^{\rm TS}$ (cm$^{-1}$)     &  1.06 & 0.41 \\\midrule
\bottomrule
\end{tabular}
\caption{PhysNet performance for the LL-PESs: Mean absolute and root
  mean squared errors (MAE and RMSE, respectively) on energies and
  forces are given for predicting the test sets of 2000 and 4000
  random structures for tropolone and PFD, respectively. The energy
  barrier for H-transfer is obtained from PhysNet ($E_{\rm B}$) and
  from \textit{ab initio} calculations ($E_{\rm B}^{\rm Ref.}$) at the
  LL of theory. The MAEs for the harmonic frequencies calculated from
  the two LL-PESs for the global minimum and the H-transfer TS as
  compared to their reference \textit{ab initio} values are also
  reported.}
\label{tab:ll}
\end{table}

\noindent
The barriers for H-transfer from the LL-PESs agree to within 0.001 and
0.01 kcal/mol with those from direct {\it ab initio} calculations at
the MP2 level which are 4.29 and 6.47~kcal/mol for tropolone and PFD,
respectively. Harmonic frequencies constitute an additional measure
for the quality of a PES around a stationary point. For the global
minimum and the transition state (TS) for H-transfer, the MAEs are
$\sim 1$ cm$^{-1}$ and $\sim 0.4$ cm$^{-1}$ for tropolone and PFD,
respectively.  Tables~\ref{sitab:harm_freq_tropo_mp2} and
\ref{sitab:harm_freq_pfd_mp2} compare the individual frequencies for
tropolone and PFD. This validates the quality of the LL-PESs compared
with reference electronic structure calculations at the MP2 level of
theory.\\

\noindent
For the present work, the main target observable is the tunneling
splitting $\Delta^{\rm H}_{\rm RPI}$ for H-transfer. Despite the
remarkable accuracy of the two LL-PESs for tropolone and PFD when
compared with the reference MP2 data, the computed tunneling
splittings overestimate the experimental results by an order of
magnitude: The H-transfer tunneling splittings on the LL-PESs are
10.94 and 0.11~cm$^{-1}$, whereas experimental
measurements\cite{tanaka:1999,daly2011microwave} give 0.974 and
0.0097211~cm$^{-1}$ for tropolone and PFD, respectively. This is a
clear indication that MP2 is not sufficient for quantitative
calculations of tunneling splittings. Hence, transfer learning to the
CCSD(T) level was explored next based on comparatively small numbers
of additional high-level calculations. The iterative data selection
procedure, which was identical for both systems, is described in
detail in the Supporting Information, with only a concise summary
provided here.\\

\begin{figure}[h!]
\centering
\includegraphics[width=1.0\textwidth]{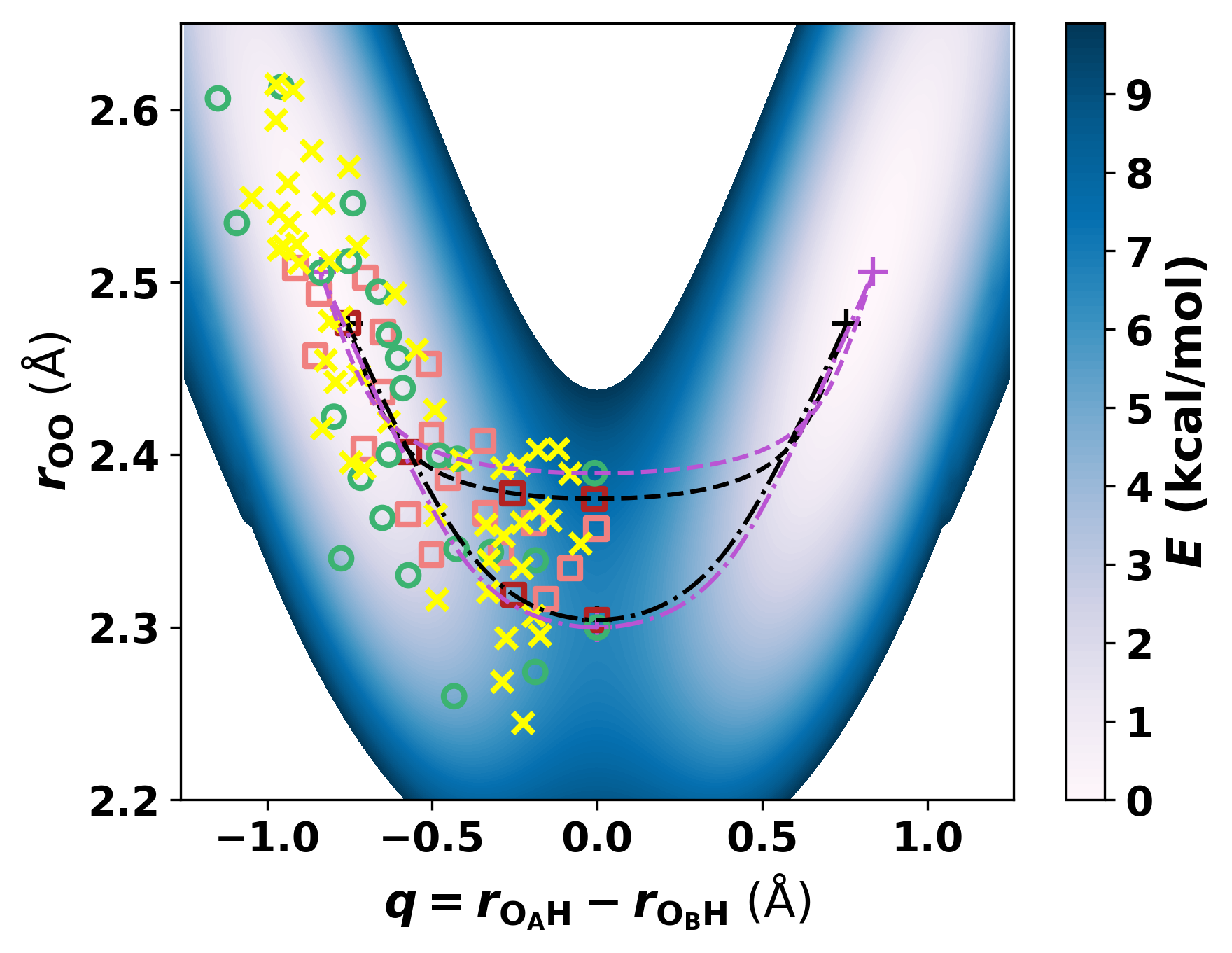}
\caption{Data sets used for TL projected onto a 2D cut through a
  representative TL-PES for tropolone spanned by the O--O distance and the
  reaction coordinate $q = r_{\rm O_AH} - r_{\rm O_BH}$. The MEPs
  (IPs) on the LL- and TL-PES are shown as dash-dotted (dashed) lines
  in black and purple, respectively. The $+$ signs mark stationary
  points on the LL- and TL-PES in black and purple, respectively. The
  TL data set is iteratively extended using active learning. Each TL
  data set is illustrated by a different marker, with the rectangles
  (brown and salmon) representing the structures employed in the first
  TL (TL$_0$), the open circles (green) are used to extend the data
  used in TL$_0$ and form the data set for TL$_1$ and the crosses
  (yellow) complement the structures used in TL$_1$ and form the data
  set for TL$_2$.}
\label{fig:rpi_tropo_2d_pes}
\end{figure}

\noindent
Based on an initial set of 25 judiciously selected structures, the
data set was iteratively extended by selecting $n_{i}$ structures
along the refined minimum energy path (MEP) and instanton path (IP),
and by selecting $n_{i,{\rm pool}}$ structures from the LL pool of
structures (see Figure~\ref{sifig:plot_surface_pool_tropo}) using
active learning. Figure \ref{fig:rpi_tropo_2d_pes} reports sets TL$_0$
to TL$_2$ for tropolone together with the MEPs (dot-dashed) and IPs
(dashed) on the LL- (black) and a representative TL-PESs (purple).
All steps involving the LL- and TL-PESs are rapid whereas the
high-level {\it ab initio} calculations are computationally demanding
($\sim 50$~h for tropolone and $\sim 100$~h for PFD). However, the
high-level \textit{ab initio} calculations for multiple structures can
be run in parallel. Due to the rather small high-level data set sizes,
TL was repeated 10 times on different splits of the data.\\

\begin{figure}[h]
\centering
\includegraphics[width=1.0\textwidth]{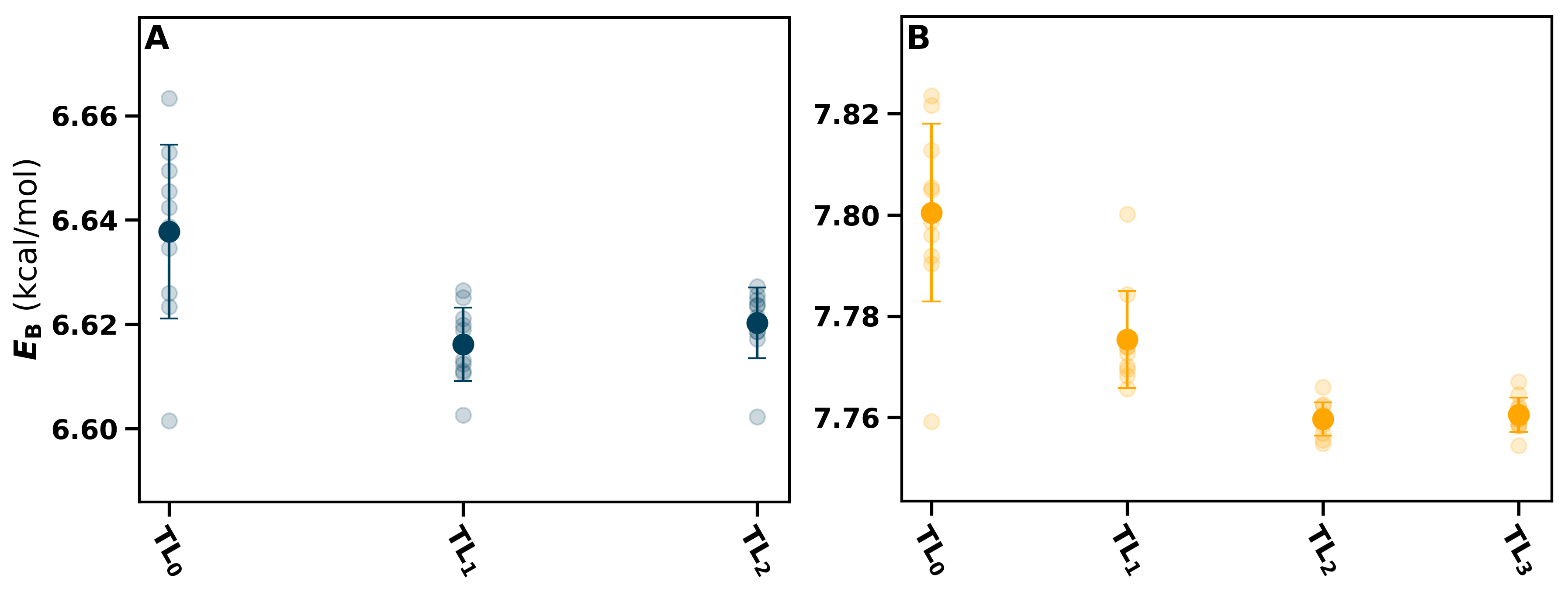}
\caption{Energy barriers $E_{\rm B}$ for H-transfer in tropolone (A)
  and PFD (B) from all ten TL-PESs (transparent circles). The
  corresponding averages (opaque circle) and standard deviations
  (error bars as $\pm\sigma$) are also indicated. An \textit{ab
    initio} optimization and subsequent determination of the energy
  barrier at the CCSD(T) level was out of reach for the two molecular
  systems.}
\label{fig:rpi_ebarrier_tropo_pfd}
\end{figure}

\noindent 
The H-transfer barrier heights for tropolone and PFD from all TLs are
shown in Figure~\ref{fig:rpi_ebarrier_tropo_pfd}. For tropolone
(Figure \ref{fig:rpi_ebarrier_tropo_pfd}A), the barrier height at the
MP2 level is 4.28~kcal/mol, which increases to 6.62~kcal/mol for
TL$_2$, \textit{i.e.}, after TL with 100 high-level data
points. Remarkably, the result with only 25 data points (TL$_0$)
appears converged as further addition of data points reduces the
uncertainty (the error bars) but not the average barrier
height. Similarly, for PFD the MP2 energy barrier of 6.46~kcal/mol
increases to 7.76~kcal/mol on the TL$_3$ PES. Again, going from TL$_0$
to TL$_3$ reduces the uncertainty and adjustments of the energy
barrier progressively decrease. It should be noted that obtaining
optimized geometries and harmonic frequencies from brute-force
electronic structure calculations at the CCSD(T)/aug-cc-pVTZ level is
not feasible for systems of the sizes studied here, but
straightforward when using the TL-PES.\cite{mm.anharmonic:2021} From
the two TL-PESs the best estimate for the barrier heights are 6.62
kcal/mol and 7.76~kcal/mol which compare with 6.63 kcal/mol and 7.76
kcal/mol from single-point CCSD(T) calculations using the optimized
and TS structures obtained from the TL$_2$ and TL$_3$ PESs for
tropolone and PFD, respectively.\\

\noindent
Previous work reported barrier heights of 6.69 kcal/mol for
tropolone\cite{khire2022enabling} from a single-point
CCSD(T)/aug-cc-pVTZ calculation using a MP2/aug-cc-pVTZ optimized
structure. Alternatively, using $\Delta$-machine learning to the
MTA-CCSD(T)/aug-cc-pVTZ level yields a barrier height of 7.18
kcal/mol.\cite{nandi2023ring} For PFD, preliminary calculations at the
CCSD(T)-F12b/aug-cc-pVDZ level found $E_{\rm B} \sim 8.0$~kcal/mol for
double H-transfer in PFD.\cite{sun2013calculations}\\

\begin{figure}[h]
\centering \includegraphics[width=1.0\textwidth]{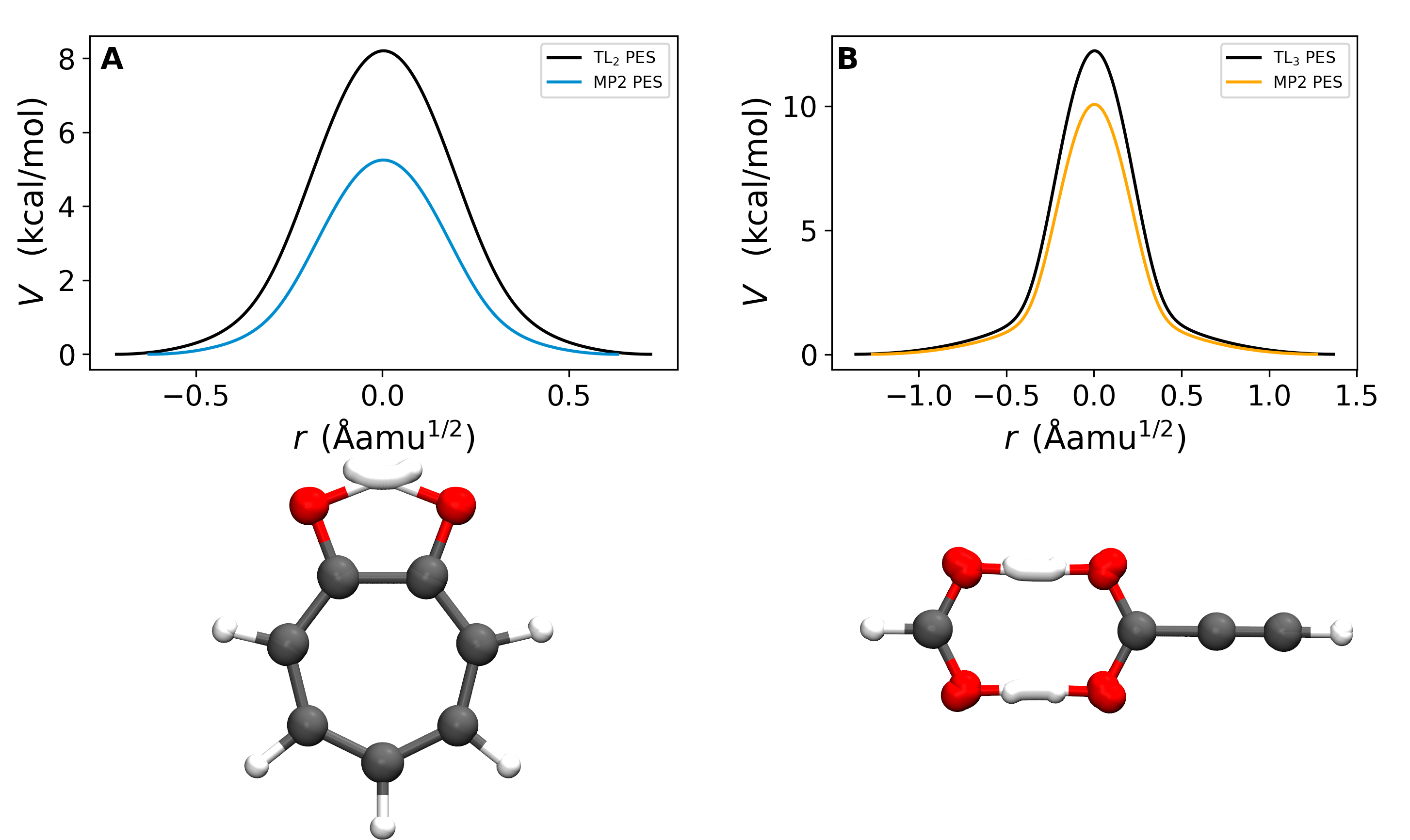}
\caption{Energy profiles (top) and corresponding instantons (bottom)
  with $N=4096$ beads from the LL- and (a representative) TL-PES for
  tropolone (A) and PFD (B). The energy profiles are given as a
  function of cumulative mass-weighted path length $r$. }
\label{fig:instpaths_tropo_pfd}
\end{figure}

\noindent
Ring polymer instanton simulations do not involve random numbers or statistical
errors which clearly distinguishes them from path-integral molecular
dynamics (PIMD)
methods\cite{Matyus2016tunnel1,Vaillant2018dimer,PIMDtunnel} or
Diffusion Monte Carlo (DMC)
simulations.\cite{kosztin1996introduction} Hence, the instanton
(once it is converged with $\tau\rightarrow\infty$ and
$N/\tau\rightarrow\infty$) is in principle uniquely determined by a
given PES. As shown in Figure \ref{fig:rpi_tropo_2d_pes} the IP
differs from the MEP.  Representative IPs on the LL- and TL-PESs for
both systems are shown in Figure~\ref{fig:instpaths_tropo_pfd} and as
a movie in the Supporting Information. For tropolone
(Figure~\ref{fig:instpaths_tropo_pfd}A), the instanton is mainly
characterized by a hydrogen motion and slight adjustments in the CC
and CO bond lengths due to single-/double-bond migration. For PFD
(Figure~\ref{fig:instpaths_tropo_pfd}B), more pronounced displacements
of the oxygen atoms, the monomer separation and wagging of the CCH
tail occur during double H-transfer.\\

\begin{figure}[h!]
\centering
\includegraphics[width=0.75\textwidth]{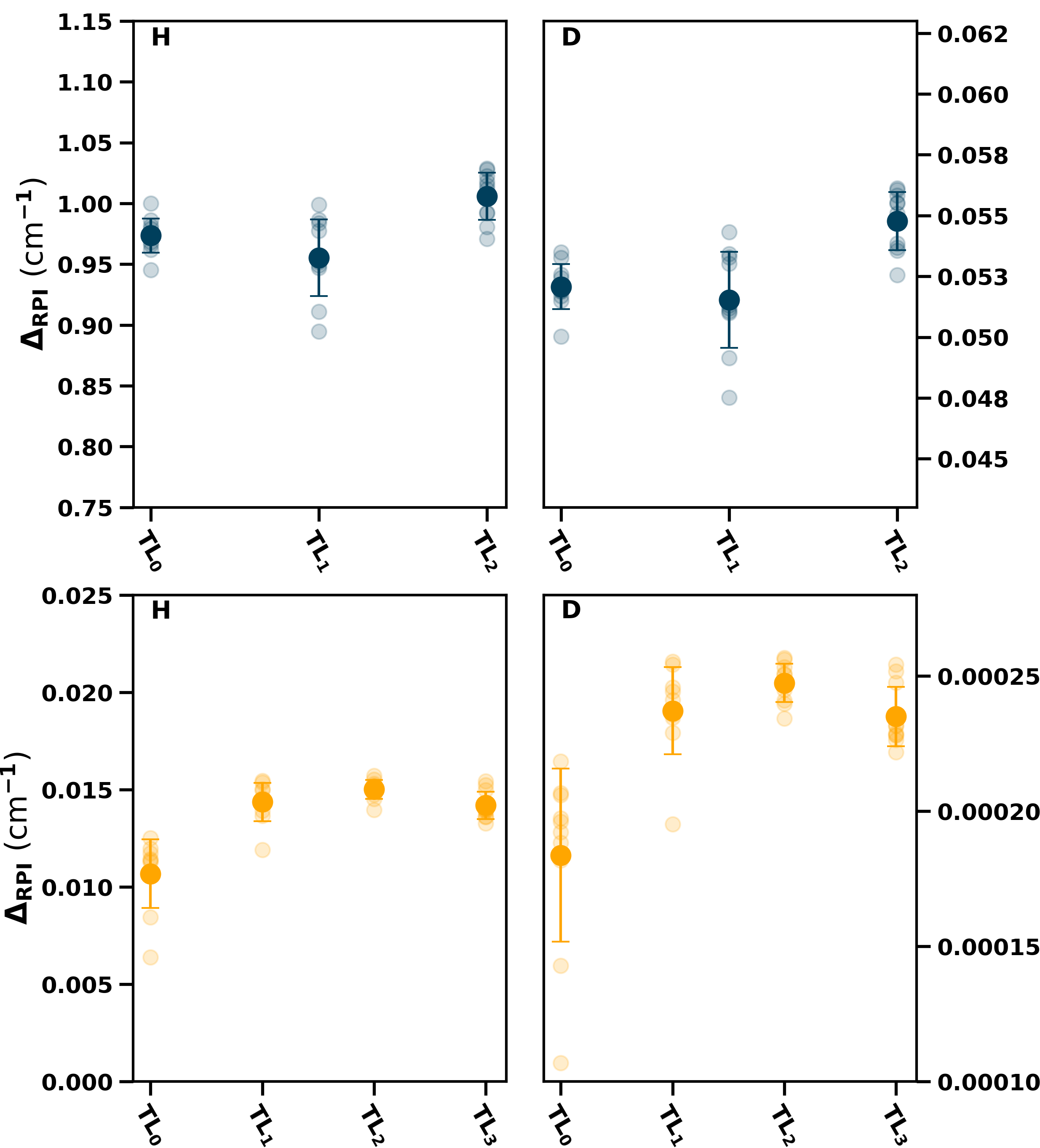}
\caption{Tunneling splittings for hydrogen (left) and deuterium
  (right) transfer in tropolone (top row) and PFD (bottom row) from
  all TL-PESs (transparent circles). The corresponding averages
  (opaque circle) and standard deviations (error bars as $\pm\sigma$)
  are also reported.}
\label{fig:rpi_splitting_tropo_pfd}
\end{figure}

\noindent
The H- and D-transfer tunneling splittings as obtained from the
different TL-PESs for tropolone and PFD are shown in
Figure~\ref{fig:rpi_splitting_tropo_pfd}. For tropolone, the instanton
result on the LL-PES of 10.94~cm$^{-1}$ is brought much closer to the
experimental result of 0.974~cm$^{-1}$ using the TL-PESs. In
particular, the experimental result is within the standard deviation
of the calculations. Note that the excellent agreement of TL$_0$ and
its rather limited fluctuation around the mean probably appears by
chance. However, it does demonstrate that even a very small number of
judiciously chosen structures (25 for TL$_0$) suffices to
significantly improve the PES and enable accurate results to be
obtained. For TL$_2$, which is arguably the most accurate PES for
tropolone, $\Delta_{\rm RPI}^{\rm H} = 1.01$ cm$^{-1}$ and
$\Delta_{\rm RPI}^{\rm D} = 0.055$ cm$^{-1}$, which both overestimate
experimental values by $\sim 5$\%, and indicate that further
improvements should be possible which is addressed further below.\\

\noindent
For PFD, gradually increasing the data set size from TL$_0$ to TL$_3$
led to smaller uncertainty and decreasing changes in $\Delta_{\rm
  RPI}$. The final transfer learning, TL$_3$, was performed to capture
the potential along the dissociation of the dimer (see
Figure~\ref{sifig:pfd_soft_mode}). As anticipated this only marginally
affected the tunneling splitting. The values of $\Delta_{\rm RPI}^{\rm
  H} = 0.014$ cm$^{-1}$ and $\Delta_{\rm RPI}^{\rm D} = 0.00024$
cm$^{-1}$ overestimate the experimentally reported splittings by $\sim
50$\% and a factor of 2, respectively. Nevertheless, they are
considerably improved over the existing computational treatments that
reported 0.63 cm$^{-1}$.\cite{sun2013calculations}\\

\begin{table}[ht]
\caption{Tunneling splittings $\Delta_{\rm RPI}^{\rm SP}$ and
  $\Delta_{\rm RPI}^{\rm DP}$ determined on the single- and
  double-precision PESs and $\Delta_{\rm PC}^{\rm DP}$ including
  perturbative corrections (PC) for tropolone and PFD. All simulations
  used the largest available TL data sets. For comparison, results
  from experiments are also reported. "n.d." is "not determined".}
    \begin{center}
    \begin{tabular}{l|lll|ll}
      \hline
    &    \multicolumn{3}{|c|}{Tropolone $\langle {\rm TL}_2\rangle$}     &    \multicolumn{2}{|c}{PFD $\langle {\rm TL}_3\rangle$}  \\
    &     \multicolumn{1}{c}{H}  &  \multicolumn{1}{c}{D}    &   H($^{13}$C)        &  \multicolumn{1}{c}{H}  & \multicolumn{1}{c}{D} \\
    \hline
    $\Delta_{\rm RPI}^{\rm SP}$ &   1.01          & 0.055  &    1.00                   &   0.0142  &  0.000235    \\
    $\Delta_{\rm RPI}^{\rm DP}$&   1.07          & 0.058  &    n.d.                    &   0.0165  &  0.000269    \\
    $\Delta_\text{PC}^{\rm DP}$ &   0.94           & 0.047  &    n.d.                   &   0.0147  &  0.000217    \\
    Expt.                                    &   0.974\cite{tanaka:1999}       &  0.051\cite{keske2006highresolution}  &    0.97\cite{keske2006highresolution}                &   0.009721\cite{daly2011microwave} &    0.0001117\cite{daly2011microwave}   \\
        \hline
  \end{tabular}
    \end{center}
\label{tab:tl_tropo_pfd_deltas} 
\end{table}

\noindent
The tunneling splittings for the two systems considered here as
determined on the largest available TL data sets are summarized in
Table~\ref{tab:tl_tropo_pfd_deltas}. For tropolone, the tunneling
splitting of the symmetric $^{13}$C isotopologue has been determined
experimentally to be 0.97~cm$^{-1}$, compared with $\Delta_{\rm
  RPI}^{\rm SP} = 1.00$~cm$^{-1}$ from the present simulations. For
the $^{18}$OH$^{18}$O isotopologue only an experimentally estimated
value $\Delta_{\rm Expt.}^{\rm ^{18}O} \sim 0.83$~cm$^{-1}$ is
available so
far.\cite{keske2006highresolution,redington2008tunnelingtropo} Hence,
the splitting was computed and yields $\Delta_{\rm RPI}^{\rm SP,
  ^{18}O} = 0.95$~cm$^{-1}$.  Due to well-understood error
cancellations, instanton theory is normally particularly reliable in
predicting isotope effects.\cite{richardson2018ring} Thus, the ratio
of calculated results can be used to obtain a new estimate of the true
tunneling splitting as $\Delta_{\rm Expt.}^{\rm ^{18}O} \sim
\Delta_{\rm Expt.}^{\rm H} \cdot \Delta_{\rm RPI}^{\rm SP, ^{18}O} /
\Delta_{\rm RPI}^{\rm SP, H} = 0.974\cdot0.95/1.01 =
0.916$~cm$^{-1}$. This value is considered an improved estimate for
$\Delta_{\rm Expt.}^{\rm ^{18}O}$ over $\Delta_{\rm Expt.}^{\rm
  ^{18}O} \sim 0.83$~cm$^{-1}$ and is a testable prediction for future
experiments from the present simulations. In all cases, the computed
splittings are larger than those reported experimentally. This is at
least partly due to inherent approximations of instanton theory.\\

\noindent
Recently, a correction scheme for RPI-based tunneling splittings using
third- and fourth-order derivatives was
proposed.\cite{richardson2023pcrpi} For cases where the higher-order
derivatives are carried out numerically, it was shown that numerically
stable applications of this correction scheme require double-precision
arithmetics for training and inference of a neural network based
PES.\cite{kaeser2023numerical} This goes beyond standard procedures
because machine learning models typically operate at 32-bit accuracy
(single-precision) to reduce the computational cost for training and
inference, and to lower memory requirements.\cite{hickmann:2020} It
was found that single-precision can have detrimental effects
particularly if higher-order derivatives of the energy are
required.\cite{kaeser2023numerical} For that reason, TL was repeated
using double-precision for the largest TL data set sizes and
$\Delta_{\rm RPI}^{\rm DP}$ was determined together with the
perturbatively corrected value $\Delta_\text{PC}^{\rm DP}$. Due to the
higher computational burden of the double-precision
training/inference, the iterative data selection procedure was
performed using single-precision.\\

\noindent
The averaged double-precision tunneling splittings $\Delta^{\rm DP}$
(averaged over 10 independent TLs) for tropolone and PFD are presented
in Table~\ref{tab:tl_tropo_pfd_deltas}. It is interesting to note that
even the uncorrected tunneling splittings $\Delta_{\rm RPI}^{\rm SP}$
and $\Delta_{\rm RPI}^{\rm DP}$ differ slightly between single- and
double-precision arithmetics. These differences are probably because
single-precision does not give the Hessians to sufficient accuracy,
leading to small errors in the fluctuation factors
$\Phi$. Comprehensive results to support this, including energy
barriers, actions $S$, fluctuation factors $\Phi$, uncorrected
tunneling splittings and correction factors $c_{\rm PC}$ for the
individual TLs are given in
Tables~\ref{sitab:tl2_ensemble_results_tropo_f32},
\ref{sitab:tl2_ensemble_results_tropo},
\ref{sitab:tl3_ensemble_results_pfd_f32} and
\ref{sitab:tl3_ensemble_results_pfd}.  \\

\noindent
For tropolone, the uncorrected ($\Delta_{\rm RPI}^{\rm DP}$) and
perturbatively corrected ($\Delta_\text{PC}^{\rm DP}$) tunneling
splittings are in excellent agreement with experiment and are the most
accurate results reported so far. Empirically, standard RPI theory for
$\Delta_{\rm RPI}$ was found to give splittings within $\sim 20$\% of
fully quantum-mechanical calculations using the same PES for typical
molecular systems, as long as the barrier height is significantly
higher than the zero-point energy along the tunneling
mode.\cite{tunnel,richardson2017full,jahr2020instanton} For
malonaldehyde\cite{mm.tlrpimalonaldehyde:2022} and tropolone, the
deviations from experiment are 16\% and 10\%, respectively, which is
within the expected range. If the results are perturbatively corrected
($\Delta_\text{PC}^{\rm DP}$), they are within 2\% to 3\% for
malonaldehyde\cite{richardson2023pcrpi,kaeser2023numerical} and
tropolone, respectively. As such, contrary to earlier
calculations\cite{nandi2023ring} that do not include perturbative
corrections, the present calculations yield the ``right answer for the
right reason''.\\

\noindent
For PFD, however, the calculated splitting for H-transfer is roughly
50\% larger than that found in the experiments, even after the
perturbative correction is applied. This is reminiscent of results for
FAD\cite{richardson2017full} for which $\Delta^{\rm H}_{\rm RPI} =
0.014$~cm$^{-1}$ was found. The calculation compares with measured
splittings between 0.011 and 0.017
cm$^{-1}$\cite{ortlieb2007proton,goroya2014high,duan:2017,li2019barrier}
(with 0.011~cm$^{-1}$ from microwave spectroscopy\cite{li2019barrier}
probably being the most reliable value), which also differs by more
than 20\%.  In search of possible explanations, additional
calculations were carried out to probe (i) limitations in the quantum
chemical methods used (ii) PES representation errors and (iii)
limitations inherent in RPI theory and the perturbative correction.\\

\noindent
Item (i) can be partly assessed by rescaling the action $S$ with a
more accurate estimation of the instanton barrier height according to
$S_{\rm new} = S \cdot\sqrt{E_{\rm inst}^{\rm new} / E_{\rm
    inst}}$. The CCSD(T)/aug-cc-pVQZ instanton barrier for PFD
optimized on a representative TL-PES is $E_{\rm B}^{\rm new} =
12.609$~kcal/mol and compares to a barrier of 12.222~kcal/mol on the
TL-PES. Rescaling yields $\Delta_{\rm RPI}$ ($\Delta_{\rm PC}$) of
0.0132 (0.0117)~cm$^{-1}$ if $c_{\rm PC} = 0.89$ from $\langle {\rm
  TL}_3\rangle$ (see Table~\ref{sitab:tl3_ensemble_results_pfd}) is
used. Although the results are closer to the measurements, they still
differ by more than 20\% in particular for the leading order
estimation. To put this into context, $S-$scaling was also applied to
malonaldehyde. Using a representative
PES\cite{mm.tlrpimalonaldehyde:2022} ($E_{\rm inst}=5.398$~kcal/mol,
$\Delta_{\rm RPI} = 24.2$~cm$^{-1}$, $\Delta_{\rm PC}=22.0$~cm$^{-1}$)
and rescaling the action using $E_{\rm inst}^{\rm new} =
5.608$~kcal/mol from CCSD(T)/aug-cc-pVQZ calculations yields
$\Delta_{\rm RPI}$ ($\Delta_{\rm PC}$) of 21.9
(19.8)~cm$^{-1}$. Comparing the results with experimental
measurements\cite{firth1991tunable} of 21.583~cm$^{-1}$ this is a
clear deterioration for $\Delta_{\rm PC}$ and shows that this simple
scaling approach can only be considered as a rough estimate of the
error. While inaccuracies in the level of theory cannot be ruled out
completely, the errors for PFD remain significantly higher than for,
\textit{e.g.}, malonaldehyde.\\

\noindent
For point (ii) - deficiencies in representing the PES - it is noted
that PhysNet was used with the same hyperparameters for both tropolone
and PFD. Training on MP2 reference data yields comparable test set
errors MAE($E$)/MAE($F$) of 0.067/0.069~kcal/mol(/\AA) for tropolone
and 0.034/0.133~kcal/mol(/\AA) for PFD (see
Table~\ref{tab:ll}). Hence, it appears unlikely that representation
errors lead to larger deviations in the tunneling splittings only for
PFD.  The quality of the PES is also influenced by the choice of
structures included.  Although the structures for both tropolone and
PFD were chosen following the same procedure and although $\Delta_{\rm
  RPI}$ is converged, it cannot be ruled out that due to the soft
intermolecular modes present in PFD, additional sampling is
required. To assess this, scans of representative soft modes and their
comparison with \textit{ab initio} CCSD(T) energies are given in
Figures~\ref{sifig:pfd_soft_mode} and \ref{sifig:pfd_dimerbend} for
PFD, which show good agreement. \\

\noindent
Finally, computed tunneling splittings can be influenced by the
semiclassical approximations underlying RPI theory, see point
(iii). Given the excellent agreement between computations and
experiment for malonaldehyde and tropolone and the inferior
performance for the hydrogen-bonded dimer PFD, supplementary
calculations were performed for FAD. Using a CCSD(T)/aug-cc-pVTZ
level of theory PES for FAD\cite{kaser2022fad} and supplementing it
with 667 additional structures along and around the instanton path
yields $\Delta_{\rm RPI}^{\rm DP} = 0.0158$~cm$^{-1}$ and
$\Delta_\text{PC}^{\rm DP} = 0.0186$~cm$^{-1}$. Again and similar to
PFD, this is consistently higher than the measured value of 0.011
cm$^{-1}$. Given that the soft modes are well captured by the PES (a
comparison of the PhysNet energies with their \textit{ab initio}
CCSD(T) energies is given in Figure~\ref{sifig:fad_softmodes}) it
seems likely that RPI theory faces challenges when applied to H-bonded
dimers, for example, due to strongly anharmonic out-of-plane
modes. This hypothesis can be tested in future work using
non-perturbative approaches such as PIMD to calculate the
tunneling splittings for a given PES.\cite{PIMDtunnel} \\

\section{Conclusion}
The present work demonstrates that combining state-of-the-art
electronic structure and machine learning techniques with RPI theory
yields H- and D-transfer tunneling splittings in very good agreement
with experiments. The calculations for tropolone are the most accurate
to date due to the high-quality PES and because they include a
perturbative correction, leading to the ``right answer for the right
reason.'' In contrast, the best previous calculation of
0.92~cm$^{-1}$, which was carried out without these corrections,
probably benefited somewhat from cancellation of errors (i) incurred
by the instanton approximation, and (ii) arising from the
fragmentation scheme for obtaining a near-CCSD(T) quality
PES.\cite{nandi2023ring} Key to the success in the present work is the
combination of transfer learning\cite{mm.tlrpimalonaldehyde:2022} and
including the perturbative correction to
RPI.\cite{richardson2023pcrpi} Importantly, brute-force calculation of
the required number of reference data (energies and forces) at the
CCSD(T) level of theory for training a machine learning PES is
unfeasible at present and for the foreseeable future. Likewise,
the Gaussian Process Regression instanton
approach\cite{laude2018ab,newGPR} would probably also not be possible
as it requires computation of a few Hessians at the CCSD(T) level.  \\

\noindent
Given the excellent performance of TL combined with RPI theory and
perturbative corrections for tunneling splittings of intramolecular
H-transfer (malonaldehyde and tropolone) it is anticipated that the
methods used in the present work can also be used to {\it
  quantitatively predict} tunneling splittings for similar
systems. This is particularly relevant because experimental
information is only available for a handful of
systems\cite{firth1991tunable,baba1999detection,redington2008infrared,murdock2010vibrational,ortlieb2007proton,duan:2017,caminati:2019,insausti2022rotational,videla2023proton,bruckhuisen2021intramolecular,tautermann2003ground}
and reliable predictions facilitate exploration of as of now
uncharacterized systems. In a yet broader context, together with
complementary experimental information such as vibrational
spectroscopic data, tunneling splittings provide valuable information
to characterize the PES around and away from the minimum energy
structure for further refinements, for example by using ``morphing''
techniques.\cite{MM.morphing:1999,MM.morphing:2024}\\

\noindent
The strategy followed here is generic and applicable to molecules for
which a few dozen very expensive {\it ab initio} calculations for
judiciously chosen geometries can be carried out. Application to yet
larger molecules necessitates advancements in either hardware
capabilities or electronic structure theory. Concerning the latter,
considerable effort is put into developing linear scaling CCSD(T)
methods, which are becoming competitive alternatives for accurate
computations of molecules exceeding 30
atoms.\cite{schutz2001low,sorathia2024improved} Alternatively,
similarly to the GEMS approach\cite{unke2024biomolecular}, machine
learning PESs can be built from molecular fragments that allow the
generation of \textit{ab initio} reference data at higher levels of
theory. Since these approximate methods are inherently difficult to
test, the framework and predictions for, \textit{e.g.}, tropolone
provided here serve as a valuable benchmark for future
applications. Finally, the present approach - with suitable
adaptations in the data selection procedure - can also be applied to
other observables that are computationally expensive to determine
(\textit{e.g.}, quantum bound states of molecules or scattering cross
sections) or be combined with other quantum dynamics methods including
wavepacket propagation or PIMD simulations.\\

\section*{Conflict of Interest}
The authors declare no conflict of interest.

\section*{Acknowledgment}
The authors gratefully acknowledge financial support from the Swiss
National Science Foundation through grants $200020\_219779$ (MM),
$200021\_215088$ (MM), the NCCR-MUST (MM and JOR), the AFOSR (MM), and
the University of Basel (MM).\\

\noindent\textsf{\textbf{Keywords:} \keywordsanie}


\setcounter{section}{0}
\setcounter{figure}{0}
\setcounter{table}{0}
\renewcommand{\thepage}{S\arabic{page}}
\renewcommand{\thetable}{S\arabic{table}}
\renewcommand{\thefigure}{S\arabic{figure}}

\clearpage
\section*{Supporting Information: Accurate Tunneling Splittings for Ever-Larger Molecules from Transfer-Learned, CCSD(T) Quality Energy Functions}

\date{\today}

\section{Computational Methods}
In the following, the RPI method to calculate tunneling splittings is
introduced, followed by a detailed overview of the construction of the
low-level (LL) and high-level (HL) PESs using PhysNet.\cite{MM.physnet:2019}

\subsection{Ring-Polymer Instanton Theory}
The RPI method offers a semiclassical approximation for computing
tunneling splittings in molecular
systems.\cite{tunnel,InstReview,hexamerprism,richardson2017full} In a
one-dimensional model, instanton theory is strongly related to the WKB
approximation.\cite{Milnikov2001} Its main advantage, however, is that
it can also be applied to multidimensional systems, in which it
locates the optimal tunneling pathway.\cite{Perspective} This pathway,
known as the instanton, is defined as an imaginary-time
$\tau\rightarrow\infty$ path connecting two degenerate wells which
minimizes the action, $S$. In computations, the path is constructed by
using an efficient ring-polymer optimization based on discretizing the
path into $N$ ring-polymer beads and taking the limit
$N\rightarrow\infty$ (typically $N \sim 1000$ is sufficient for
convergence). The potential $U_N$ of a ring-polymer is given by
\begin{align}
  \label{eq:rpi_polymer_potential}
    U_N(\bm{x};\beta) = \sum_{i=1}^N V(x_i) +
    \frac{1}{2(\beta_N\hbar)^2}\sum_{i=1}^N (x_{i+1} - x_i)^2 \equiv
    \frac{S({\bf x})}{\beta_N \hbar}
\end{align}
with $\bm{x} = \left( \bm{x}_1, \dots, \bm{x}_N\right)$ being the
mass-scaled coordinates of the beads, $\beta = \frac{1}{k_b T}$ and
$\beta_N = \beta/N$. The first term in
Equation~\ref{eq:rpi_polymer_potential} corresponds to the sum over
all single bead potentials and the second term represents the harmonic
springs with frequency $1/(\beta_N\hbar)$ that connect the different
beads. As the potential $U_N$ and the action $S$ are related by
division with $(\beta_N\hbar)$, $S$ is determined by the distance
between neighbouring beads as well as the potential energy of each
bead. Therefore, the action $S$ contains information
\emph{along} the instanton path (IP). A detailed description of the
method is provided, \textit{e.g.}, in References~\citenum{tunnel} and
\citenum{InstReview}.\\

\noindent
Once the IP has been located, fluctuations around the path are
computed to second order and the information is combined into the term
$\Phi$, \textit{i.e.}, this is based on information \emph{around} the
IP. This requires Hessians at each of the beads. The prediction
for the leading order tunneling splitting (in a double-well
system) is
\begin{align}
    \Delta_{\rm RPI} = \frac{2\hbar}{\Phi} \sqrt\frac{S}{2\pi\hbar} \,
    \mathrm{e}^{-S/\hbar}.
\label{sieq:rpi_splitting} 
\end{align}
Because $S$ appears in the exponent it is particularly important to
determine this quantity with high accuracy. A limitation of the RPI
method for determining tunneling splittings is that fluctuations
around the instanton are treated harmonically (\textit{i.e.}, the
Hessians enter via $\Phi$). To a good approximation, it is expected
that this captures the dominant tunneling contribution, except for
cases in which anharmonic effects perpendicular to the instanton are
significant or where the barrier is low. For this reason, Richardson
and coworkers recently developed a perturbatively corrected RPI
theory\cite{richardson2023pcrpi}, that accounts for additional
anharmonicity by including information from the third and fourth
derivatives of the potential along the instanton. The tunneling
splitting obtained from perturbatively corrected RPI theory is denoted
as $\Delta_{\rm PC}$ in the following, which, in practice is obtained
from scaling $\Delta_{\rm RPI}$ with a correction factor $c_{\rm PC}$
following $\Delta_{\rm PC} = c_{\rm PC}\cdot \Delta_{\rm RPI}$.\\

\noindent
All leading order splitting calculations in the present work were
carried out with $N=4096$ beads and at an effective temperature of
$T_{\rm eff} = \hbar/k_{\rm B} \tau = 25$~K. The numerical convergence
of the leading order tunneling splittings is reported in
Tables~\ref{sitab:convergence_tropo}(a) and
\ref{sitab:convergence_pfd}(a).  The perturbatively corrected
instanton calculations were carried out at $T_{\rm eff} = 50$~K and
with $N=512$ beads. The convergence sets are
reported in Tables~\ref{sitab:convergence_tropo}(b) and
\ref{sitab:convergence_pfd}(b) and show that $c_{\rm PC}$ converges
significantly more rapidly than the leading order term.

\subsection{Low-Level PESs}
\begin{figure}[h]
\centering
\includegraphics[width=0.9\textwidth]{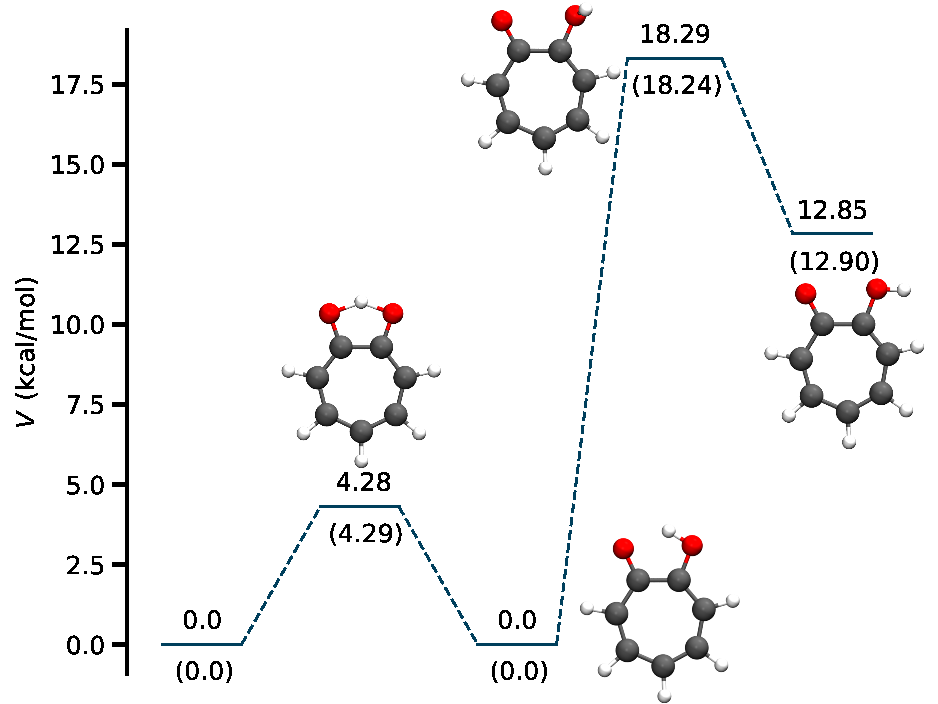}
\caption{Schematic of the LL-PES for tropolone covering all
  configurations. The energies in brackets correspond to \textit{ab
    initio} values.}
\label{sifig:pes_tropo_ll}
\end{figure}

\begin{figure}[h]
\centering
\includegraphics[width=0.9\textwidth]{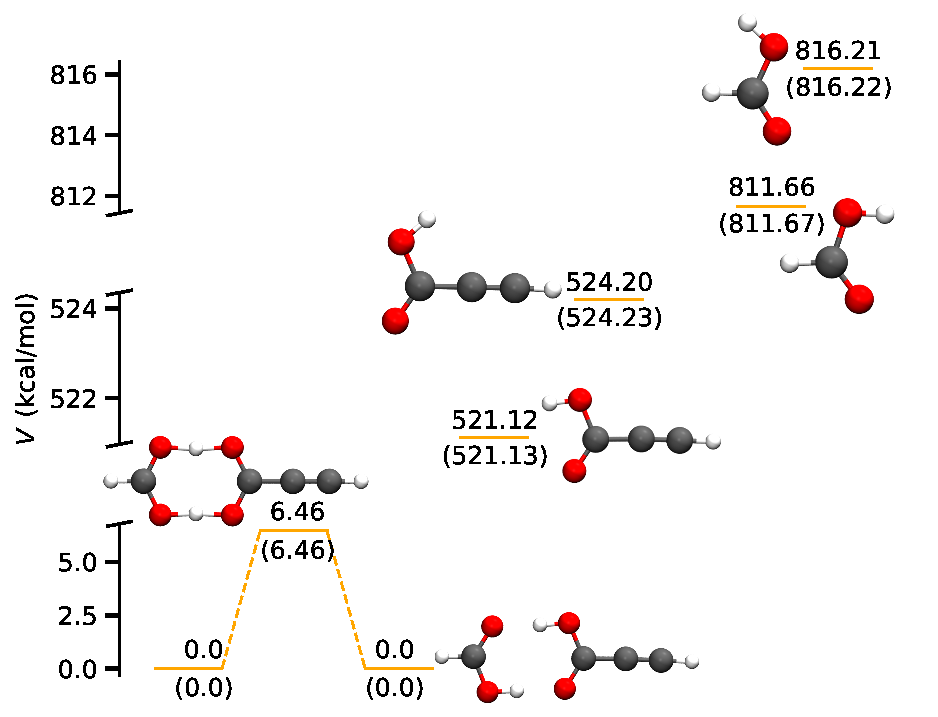}
\caption{Schematic of the LL-PES for PFD covering all
  configurations. The energies in brackets correspond to \textit{ab
    initio} values. The energies correspond to atomization energies
  (\textit{i.e.}, the atomic energies have been subtracted) and the energy of
  the minimum dimer structure serves as reference.}
\label{sifig:pes_PFD_ll}
\end{figure}

The data generation to construct PESs for tropolone and PFD (including
minima and transition states (TS) for hydrogen transfer) followed
similar strategies. While for tropolone sampling two conformers and
the corresponding TSs was sufficient, the PES for PFD involves
multiple conformational substates and dissociation products including
\textit{cis-} and \textit{trans-}formic acid, \textit{cis-} and
\textit{trans-}propiolic acid. The conformations and their relative
energies for both systems are illustrated in
Figures~\ref{sifig:pes_tropo_ll} and \ref{sifig:pes_PFD_ll}. Including
monomers and dissociation products for PFD is not anticipated to
improve the tunneling splitting accuracy. However, accounting for
these states is essential for a \textit{global} PES and desirable
future investigations.\\

\noindent
First, normal mode vectors and force constants for all vibrations were
obtained at the MP2/aug-cc-pVDZ level of theory using
Gaussian.\cite{gaussian16} This information was used to carry out
normal mode sampling\cite{smith2017ani} at temperatures $T = [100,
  300, 500, 100, 1500, 2000]$~K to yield 13000 and 15000 structures
for tropolone and PFD, respectively. For tropolone 22 additional
structures were obtained along the CH-stretching mode to give a total
of 13022 structures. For PFD additional structures were generated by
running $NVT$ MD simulations at 1000~K for \textit{cis}- and
\textit{trans}-formic acid, \textit{cis}- and \textit{trans}-propiolic
acid and PFD (9000 structures in total) using the semiempirical
GFN2-xTB method\cite{bannwarth2019gfn2} as available in the atomic
simulation environment.\cite{larsen2017atomic} The region around the
TS as well as the non-hydrogen bonded complex of the monomers (van der
Waals complex) constituting PFD were sampled using a biasing harmonic
potential to control structural sampling (7000 structures in
total). Thus, the initial data set contained 31000 structures for PFD
and corresponding monomers. For all these structures, energies, forces
and dipole moments were determined at the MP2/cc-pVTZ level of theory
using MOLPRO.\cite{MOLPRO} For both systems two independent PhysNet
models were trained.\\

\noindent
The PES pairs were used to carry out adaptive sampling
simulations.\cite{csanyi2004learn} For tropolone these were two rounds
of $(NVT)$ simulations at 1000 and 2000~K, respectively. For PFD
adaptive sampling was run at 1000~K for the six stationary points on
the PES (see Figure~\ref{sifig:pes_PFD_ll}) and the van der Waals
complex of the (\textit{trans}-) formic acid and propiolic acid
monomers. Structures were saved if the predictions of the two PESs
differed by more than 0.5~kcal/mol. All adaptive sampling MD
simulations were carried out using the Langevin scheme with a time
step of 0.1~fs as implemented in ASE.\cite{larsen2017atomic} This
resulted in an additional 5619 structures for tropolone and 9013
structures for PFD (most of which cover the van der Waals complex,
\textit{i.e.}, 6300), respectively.\\

\noindent
For both systems (tropolone and PFD with substates) the
configurational space of the PES was further sampled using diffusion
Monte Carlo simulations (DMC),\cite{kosztin1996introduction} from
which 1356 structures were extracted for tropolone whereas for PFD 277
additional structures were obtained. All DMC simulations employ a GPU
implementation of the DMC algorithm for PhysNet and were run with 300
walkers, a step size of 5 a.u. and for a total of 60000
steps. Unphysical structures were detected and added to the training
dataset if the predicted energy was lower than the energy of the
minimum energy structure.\\

\noindent
Finally, for both systems two independent PESs were retrained based
on a total of 19997 structures for tropolone and 40290 structures for
PFD. Using the two PES pairs another round of adaptive sampling and DMC
simulations was carried out. However, no further deficient structures
were detected.\\

\subsection{High-Level PESs / Data Selection Procedure}
\label{sec:methods_highlevel}

\begin{figure}[h]
\centering
\includegraphics[width=0.9\textwidth]{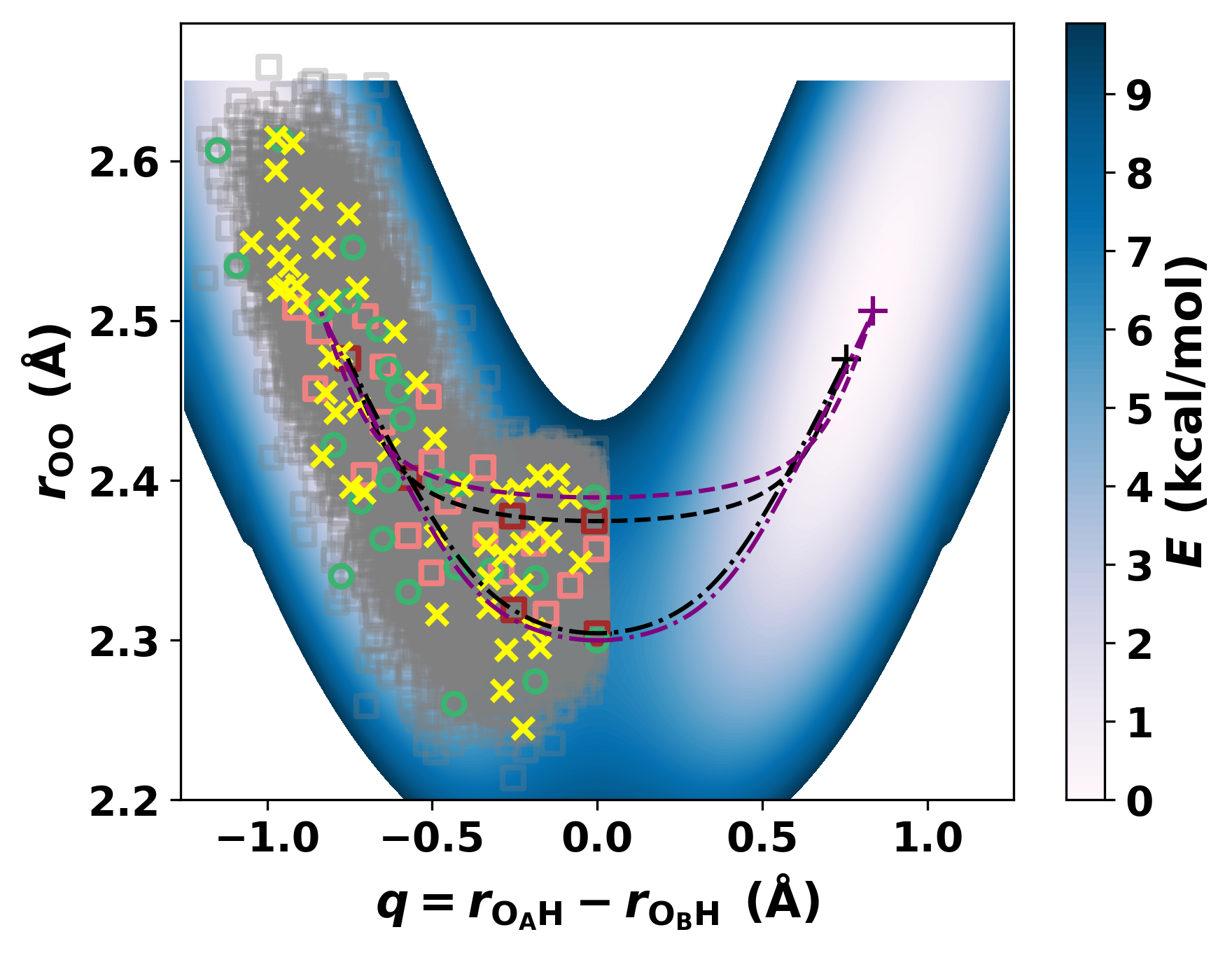}
\caption{Data sets used for TL projected onto a 2D cut through the
  final TL PES for tropolone spanned by the O--O distance and the
  reaction coordinate $q = r_{\rm O_AH} - r_{\rm O_BH}$. The MEPs
  (IPs) on the low-level and TL PES are shown as dash-dotted (dashed)
  lines in black and purple, respectively. The $+$ signs mark
  stationary points on the low-level and TL PES in black and purple,
  respectively. The TL data set is iteratively extended based on the
  performance of the TL PESs. Each TL data set is illustrated by a
  different marker, with the rectangles (brown and salmon)
  representing the structures employed in the first TL (TL$_1$), the
  open circles (green) are used to extend the data used in TL$_1$ and
  for the data set for TL$_2$ and the crosses (yellow $\times$
  symbols) complement the structures used in TL$_2$ and for the data
  set for TL$_3$. The grey symbols illustrate the coverage of the
  "structure pool" from which appropriate structures are selected.}
\label{sifig:plot_surface_pool_tropo}
\end{figure}

\noindent
It has been recently demonstrated that transfer learning a LL-PES to a
higher level of electronic structure theory can be accomplished
efficiently by judiciously choosing reference structures for which
computationally expensive HL energies, forces and dipole moments are
determined.\cite{mm.tlrpimalonaldehyde:2022} This was shown for
malonaldehyde and it was found that transfer learning using 25 to 100
algorithmically selected structures and corresponding CCSD(T) data
performs on par with a PES that was transfer-learned using $\sim 900$
structures. Following the road-map towards high-level of theory PESs
presented in Reference~\citenum{mm.tlrpimalonaldehyde:2022},
high-level information was chosen for tropolone and
PFD. Figure~\ref{sifig:plot_surface_pool_tropo} illustrates the data
selection procedure for tropolone, which, however, is representative
for both systems. Note that it is not necessary to sample both
potential wells since PhysNet handles molecular symmetry by
construction.\\

\begin{enumerate}
    \item The MEP (black dash-dotted) and the IP (black dashed) are
      determined on the LL-PES. From these, a total of six
      structures (brown rectangles), including the minimum structure,
      the TS structure, the structure of the IP that has the highest
      potential energy (\textit{i.e.}, at $q=0$) and three additional
      structures that are evenly distributed are manually selected.
    \item Without additional {\it a priori} information it is
      advantageous to generate an initial pool of structures, that
      sample all relevant regions of the PES, to select from (note
      that at this stage no \textit{ab initio} calculations are
      performed - the structures are sampled using MD, normal mode
      sampling on the low-level PES or are available as part of the
      LL-PES reference data). The coverage of the "structure pool"
      from which appropriate structures are selected is shown in
      Figure~\ref{sifig:plot_surface_pool_tropo}. Structures are
      selected from the pool such that their structural RMSD is
      maximized with respect to the existing set of six
      structures. These 19 structures are illustrated as salmon
      rectangles in Figure~\ref{fig:rpi_tropo_2d_pes}.
    \item The 25 structures selected above (rectangles) form the data
      set for the first TL, which is referred to as TL$_0$ in the
      following. For these, CCSD(T) energies, forces and dipole
      moments are calculated using MOLPRO, which are used to
      transfer-learn the NN.
    \item Next, the MEP (purple dash-dotted) and the IP (purple
      dashed) are determined on the TL$_0$ PES. From these, a total of
      five structures including the new minimum structure, the new TS
      structure and the new structure of the IP that has the highest
      potential energy and two evenly distributed structures are
      manually selected.
    \item The five structures above are augmented with 20 structures
      that are chosen with active learning (\textit{i.e.}, two
      independent TL PESs are used to predict the pool of structures
      and the structures with the highest $\Delta E$ are selected).
    \item For the 25 structures selected above (green open circles),
      CCSD(T) energies, forces and dipole moments are calculated using
      MOLPRO. They, together with the 25 structures from TL$_0$ form
      the data set for TL$_1$. The LL-PES is transfer-learned with the
      50 structures selected above (circles and rectangles).
    \item The MEP and the IP are determined on the TL$_1$ PES (not
      shown in Figure~\ref{fig:rpi_tropo_2d_pes} because they are the
      same as for the TL$_0$ PES). If the minimum structure, the TS
      structure and the IP structure with the highest potential energy
      differ from the structures selected in the previous iteration,
      then they are selected again. This was not found to be necessary
      here.
    \item Active learning is again employed to select the $N$ (here
      $N=50$) structures with the highest $\Delta E$ form the pool.
    \item For the $N$ structures selected above (yellow $\times$
      symbols), CCSD(T) energies, forces and dipole moments are
      calculated using MOLPRO. They, together with the 50 structures
      from TL$_1$, form the data set for TL$_2$.
    \item Repeat steps 7. to 9. until the observable(s) of interest
      converge. This closes the loop of the recommended procedure to
      obtain spectroscopically accurate PESs.
\end{enumerate}

\noindent
The procedure outlined above proved to be effective but may need to be
adapted for other systems, \textit{e.g.} selecting a smaller or larger
number of structures in each iteration. The TL data sets for tropolone
contained 25/50/100 structures for TL$_0$/TL$_1$/TL$_2$, while for PFD
a fourth TL was carried out to correct the PES in the dissociation
limit of the two monomers. Thus, the TL data sets for PFD contained
25/50/100/(106 PFD + 300 monomer structures) for
TL$_0$/TL$_1$/TL$_2$/TL$_3$. Due to the rather limited data set sizes,
TL was repeated ten times using different data splits for each system
and TL data set size.\\

\noindent
In the present case and on the computers used, the determination of
\textit{ab initio} energy/force/dipole moment for a single structure
takes roughly 50~h for tropolone and 100~h for PFD. The high-level
\textit{ab initio} calculations were performed at CCSD(T) level with
aug-cc-pVTZ basis set for PFD while for tropolone a slightly smaller
basis set was used (aug-cc-pVTZ basis on the atoms that are involved
in the hydrogen transfer reaction (\textit{i.e.}, transferring H-atom
and both O-atoms), while cc-pVTZ basis sets represent the rest of the
molecule.) This was necessary because of the enormous memory
requirements of such calculations.\\

\clearpage
\section{Results}
\subsection{Low-level PES}

\begin{figure}[h!]
\centering
\includegraphics[width=0.9\textwidth]{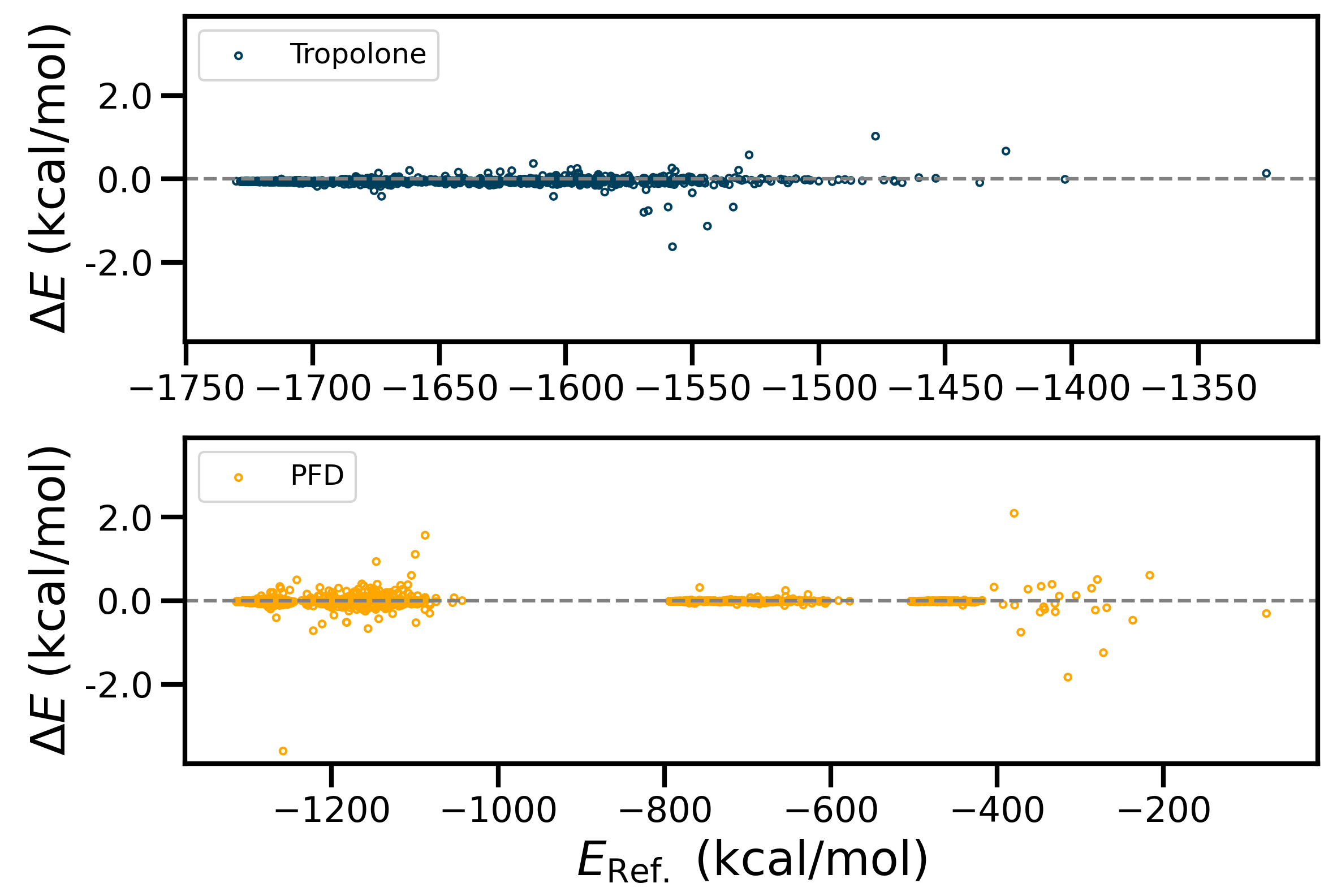}
\caption{Performance of the LL-PESs on a hold-out test set for
  tropolone and PFD containing roughly 2000 and 4000 samples,
  respectively. Here, $\Delta = E_{\rm Ref.} - E_{\rm PhysNet}$ and
  contains all but a single outlier from the PFD model (that has
  $\Delta\approx8$~kcal/mol at -200~kcal/mol) which was omitted for
  better visibility of the absolute errors. The outlier structure
  exhibits a very distorted configuration and is typically generated
  at an early stage of the adaptive sampling when the global PES is
  still insufficiently sampled.  The tropolone data set covers both
  conformers while the data set for PFD contains the dimer, and both
  monomers in \textit{cis-} and \textit{trans-}conformations. PFD,
  propiolic acid and formic acid configurations are represented by the
  point clouds at $-1200$, $-700$ and $-500$~kcal/mol, respectively.}
\label{fig:ll_corr}
\end{figure}

\begin{table}[h]
\begin{tabular}{rrrrr}
& \textbf{Min} & & \textbf{TS} & \\\toprule
\textbf{Mode} & \textbf{PhysNet} & \textbf{MP2} & \textbf{PhysNet} & \textbf{MP2}\\\midrule
1	&	116.4	&	115.2	&	1223.0i &	1225.5i	\\
2	&	183.0	&	179.4	&	155.4	&	153.0	\\
3	&	352.0	&	352.2	&	180.6	&	177.5	\\
4	&	366.0	&	366.2	&	362.2	&	361.5	\\
5	&	375.0	&	371.8	&	390.8	&	387.2	\\
6	&	399.2	&	395.3	&	400.5	&	399.9	\\
7	&	443.1	&	442.8	&	409.4	&	408.9	\\
8	&	546.5	&	546.9	&	565.2	&	565.3	\\
9	&	581.2	&	578.8	&	588.4	&	583.4	\\
10	&	701.1	&	701.6	&	716.2	&	716.8	\\
11	&	723.7	&	722.0	&	745.3	&	745.5	\\
12	&	752.6	&	753.4	&	749.2	&	748.0	\\
13	&	770.3	&	767.2	&	761.6	&	760.2	\\
14	&	850.4	&	850.0	&	772.0	&	773.0	\\
15	&	872.9	&	872.4	&	867.0	&	866.0	\\
16	&	889.4	&	890.0	&	886.7	&	887.1	\\
17	&	937.0	&	940.4	&	941.7	&	945.9	\\
18	&	978.2	&	979.3	&	978.8	&	979.3	\\
19	&	984.7	&	983.1	&	994.1	&	990.4	\\
20	&	1005.6	&	1004.8	&	994.7	&	995.7	\\
21	&	1080.8	&	1081.9	&	1103.8	&	1104.3	\\
22	&	1236.8	&	1238.1	&	1241.0	&	1242.3	\\
23	&	1243.4	&	1243.7	&	1244.5	&	1244.5	\\
24	&	1277.9	&	1278.6	&	1265.1	&	1266.3	\\
25	&	1331.3	&	1331.9	&	1296.8	&	1297.3	\\
26	&	1357.3	&	1357.7	&	1411.1	&	1411.3	\\
27	&	1448.9	&	1448.9	&	1418.7	&	1419.1	\\
28	&	1473.4	&	1474.0	&	1459.2	&	1459.0	\\
29	&	1526.7	&	1526.4	&	1506.5	&	1506.2	\\
30	&	1581.6	&	1580.7	&	1584.5	&	1584.6	\\
31	&	1617.1	&	1617.6	&	1634.5	&	1635.0	\\
32	&	1646.1	&	1645.6	&	1690.2	&	1690.2	\\
33	&	1693.3	&	1693.3	&	1749.5	&	1748.7	\\
34	&	3183.3	&	3183.8	&	2101.0	&	2100.9	\\
35	&	3192.7	&	3193.4	&	3184.7	&	3184.3	\\
36	&	3207.9	&	3208.7	&	3191.1	&	3191.1	\\
37	&	3210.3	&	3211.3	&	3212.3	&	3212.4	\\
38	&	3221.9	&	3221.9	&	3215.4	&	3215.2	\\
39	&	3302.0	&	3302.0	&	3221.3	&	3221.3	\\ \bottomrule
\end{tabular}
\caption{Harmonic frequencies for tropolone at the global minimum and
  the TS structure as calculated from the PhysNet PES and at the MP2/cc-pVTZ
  level of theory.}
\label{sitab:harm_freq_tropo_mp2}
\end{table}

\begin{table}[h]
\begin{tabular}{rrrrr}
& \textbf{Min} & & \textbf{TS} & \\\toprule
\textbf{Mode} & \textbf{PhysNet} & \textbf{MP2} & \textbf{PhysNet} & \textbf{MP2}\\\midrule
1	&	52.3	&	52.5	&	1244.4i	&	1242.2i	\\
2	&	73.9	&	73.9	&	71.0	&	71.3	\\
3	&	116.7	&	117.0	&	74.8	&	74.7	\\
4	&	176.2	&	175.9	&	130.5	&	130.6	\\
5	&	205.9	&	205.6	&	239.0	&	239.1	\\
6	&	223.7	&	224.4	&	255.2	&	255.9	\\
7	&	270.8	&	271.1	&	297.1	&	297.0	\\
8	&	283.2	&	283.9	&	414.7	&	414.3	\\
9	&	573.5	&	573.4	&	481.2	&	481.7	\\
10	&	607.1	&	607.4	&	666.7	&	667.8	\\
11	&	670.3	&	671.1	&	671.3	&	671.7	\\
12	&	673.0	&	673.0	&	690.3	&	690.5	\\
13	&	704.9	&	704.7	&	699.0	&	698.7	\\
14	&	773.6	&	773.8	&	763.0	&	763.0	\\
15	&	876.7	&	876.6	&	787.0	&	787.7	\\
16	&	976.6	&	977.3	&	924.9	&	924.7	\\
17	&	1005.2	&	1006.0	&	1090.4	&	1090.8	\\
18	&	1109.7	&	1109.8	&	1323.1	&	1323.3	\\
19	&	1269.7	&	1269.6	&	1355.3	&	1354.7	\\
20	&	1323.5	&	1323.4	&	1363.2	&	1363.6	\\
21	&	1418.8	&	1418.7	&	1421.5	&	1421.7	\\
22	&	1466.6	&	1466.5	&	1425.5	&	1425.6	\\
23	&	1504.9	&	1505.1	&	1465.9	&	1465.7	\\
24	&	1715.1	&	1715.5	&	1586.9	&	1587.0	\\
25	&	1779.8	&	1779.4	&	1697.4	&	1697.3	\\
26	&	2157.8	&	2158.0	&	1743.4	&	1742.8	\\
27	&	3095.7	&	3097.4	&	1780.4	&	1780.8	\\
28	&	3136.1	&	3135.7	&	2160.6	&	2161.3	\\
29	&	3236.5	&	3237.4	&	3152.9	&	3152.6	\\
30	&	3493.9	&	3494.3	&	3493.9	&	3494.7	\\
\bottomrule
\end{tabular}
\caption{Harmonic frequencies for PFD at the global minimum and the TS
  structure as calculated from the PhysNet PES and at the MP2/cc-pVTZ level of
  theory.}
\label{sitab:harm_freq_pfd_mp2}
\end{table}

\clearpage
\subsection{High-Level PES}
\begin{figure}[h]
\centering
\includegraphics[width=1.0\textwidth]{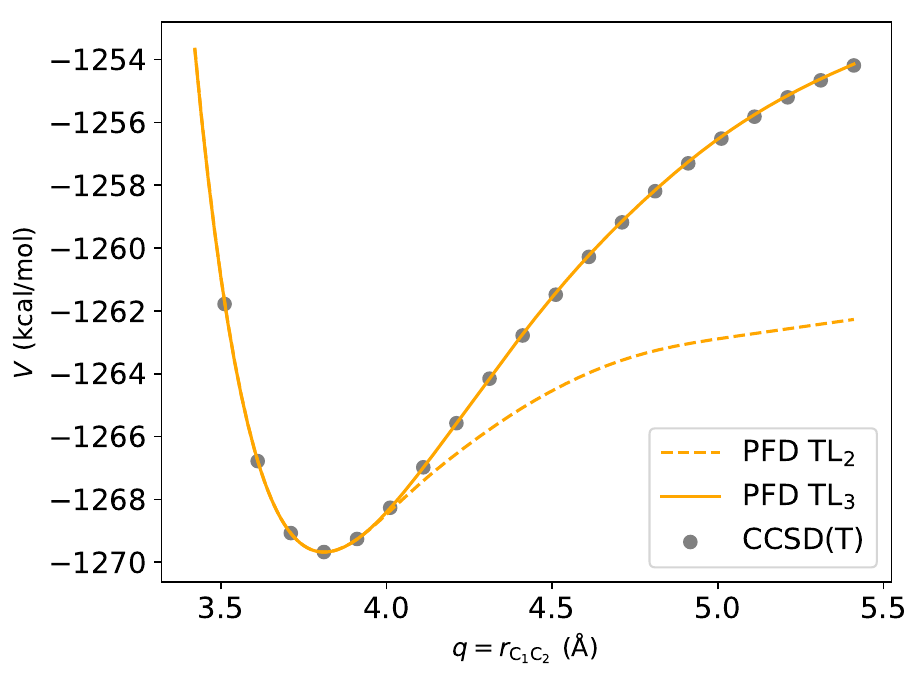}
\caption{The dissociation of PFD predicted by PhysNet (TL$_2$ and
  TL$_3$) and compared to \textit{ab initio} CCSD(T) values. TL$_3$
  includes structures along the dimer dissociation to correct the
  asymptotic behaviour. Clearly, TL$_3$ captures this correctly.}
\label{sifig:pfd_soft_mode}
\end{figure}

\begin{figure}[h]
\centering
\includegraphics[width=1.0\textwidth]{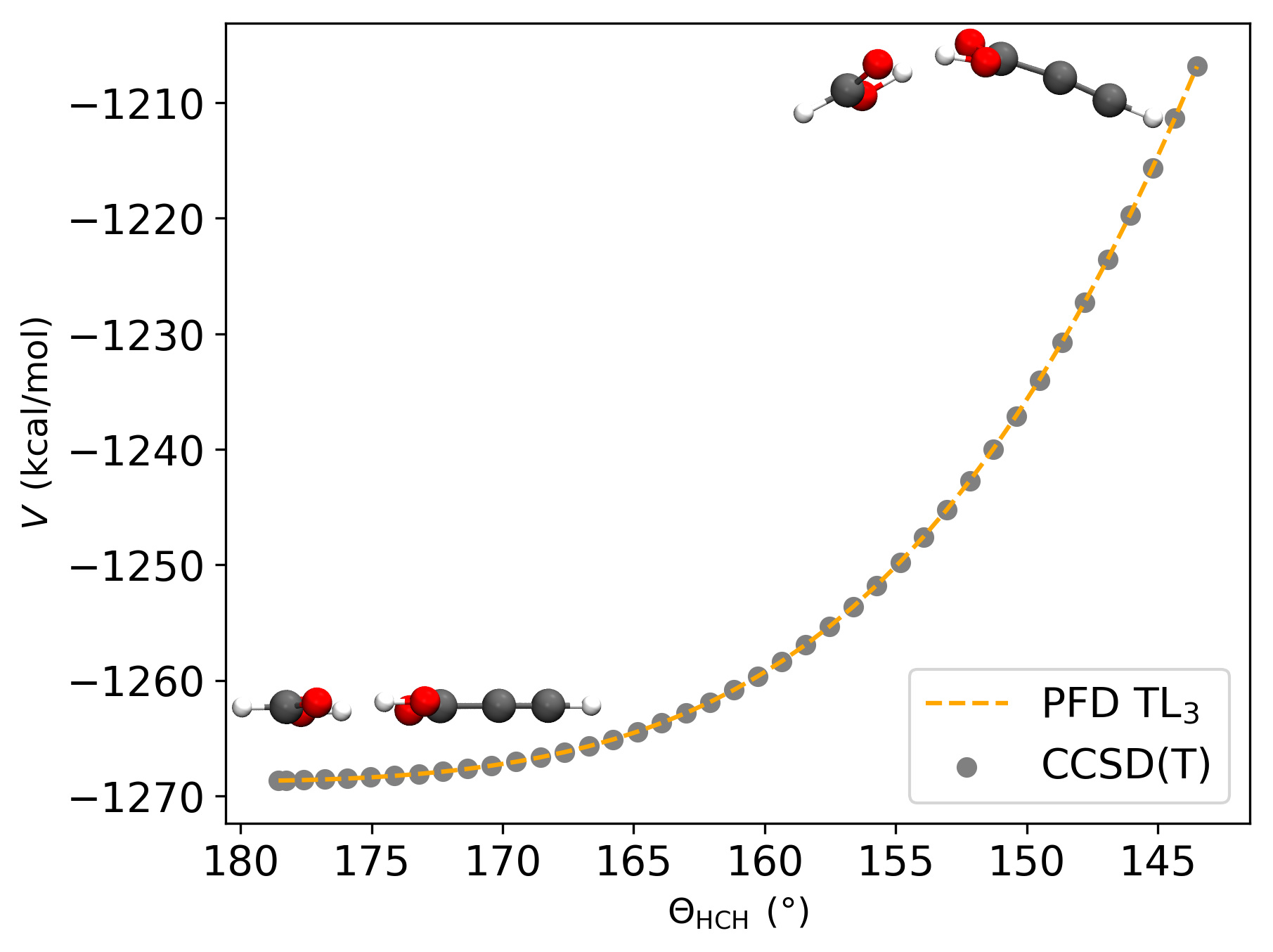}
\caption{The low-frequency out-of-plane puckering mode of PFD predicted by
  PhysNet (TL$_3$) and compared to \textit{ab initio}
  CCSD(T)/aug-cc-pVTZ values. Clearly, TL$_3$ captures this
  correctly.}
\label{sifig:pfd_dimerbend}
\end{figure}

\begin{figure}[h]
\centering
\includegraphics[width=1.0\textwidth]{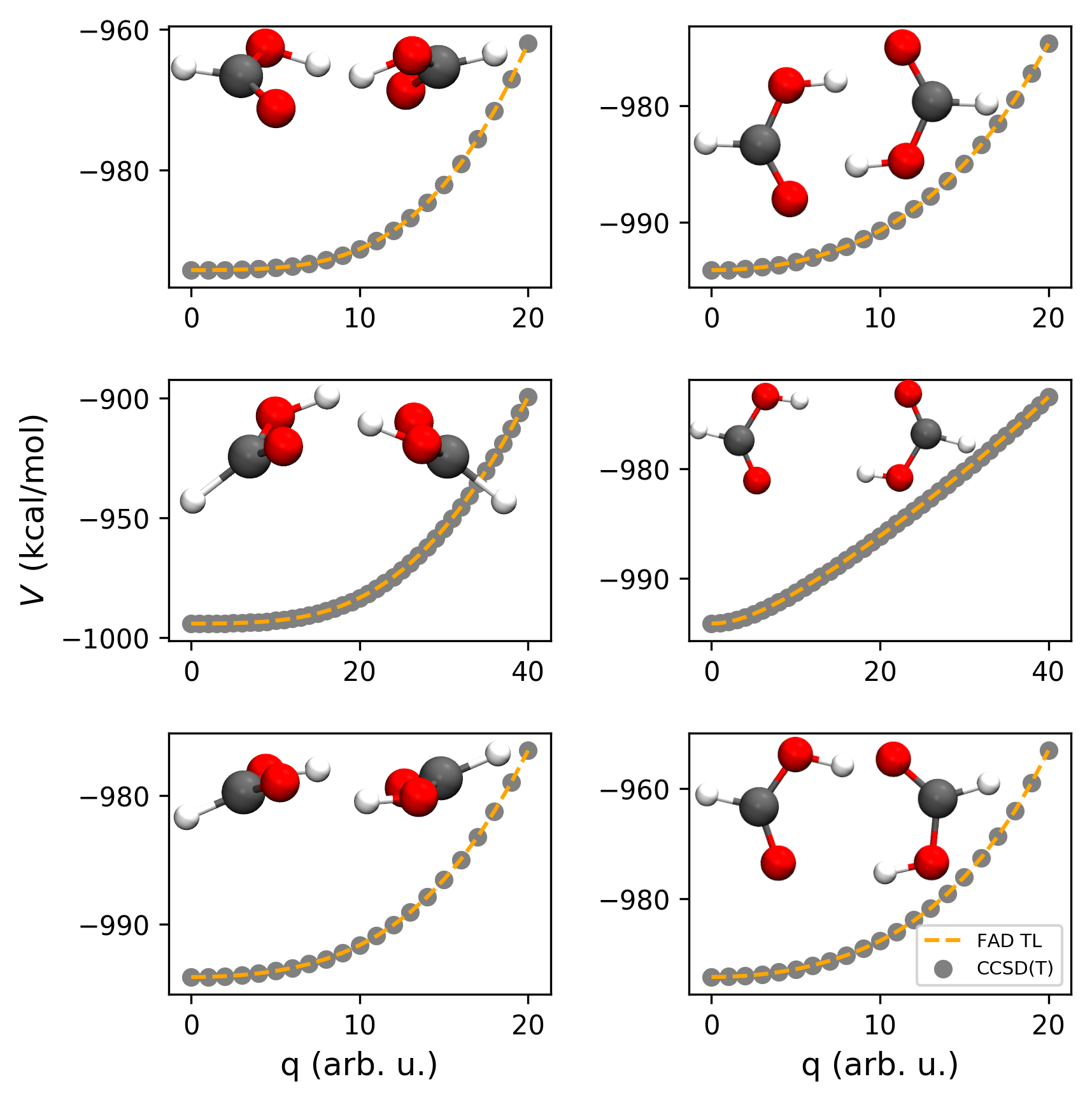}
\caption{Comparison of PhysNet and direct CCSD(T) energies for
  structures along the low-frequency modes of FAD. $q=0$ corresponds
  to the (planar) optimized structure and for increasing $q$ the
  structure is further displaced along the corresponding normal
  modes.}
\label{sifig:fad_softmodes}
\end{figure}

\clearpage
\section{Convergence of the RPI Calculations}
The section contains the convergence of the numerical results for
tropolone and PFD, respectively.

\subsection{Tropolone}

\begin{table}[h]
    \begin{subtable}[c]{0.45\textwidth}
        \centering
\begin{tabular}{rrrr}\toprule
 & \multicolumn{3}{c}{$T = \hbar/(k_B \tau)$}\\\cmidrule(lr){2-4}
 $N$ & 50~K & 25~K & 12.5~K \\\midrule
128	&	0.984	&		&	\\
256	&	1.057	&	0.967	&	\\
512	&	1.078	&	1.042	&	0.950 \\
1024	&	1.083	&	1.063	&	1.029 \\
2048	&	1.084	&	1.068	&	1.051 \\
4096	&	1.085	&	1.070	&	1.057 \\\bottomrule
\end{tabular}
\caption{Leading order tunneling splitting $\Delta_{\rm RPI}^{\rm DP}$ in
  cm$^{-1}$.}
    \end{subtable}
    \hfill
    \begin{subtable}[c]{0.45\textwidth}
        \centering
\begin{tabular}{rr}\toprule
 & $T = \hbar/(k_B \tau)$\\\cmidrule(lr){2-2}
 $N$ & 50~K  \\\midrule
128	&	0.946  \\
256	&	0.912  \\
512	&	0.912 \\\bottomrule
\end{tabular}
\caption{Perturbative correction factor $c_{\rm PC}$ (unitless).}
     \end{subtable}
\caption{Numerical convergence for hydrogen transfer in tropolone on a
  representative double-precision TL PES.}
\label{sitab:convergence_tropo}
\end{table}

\begin{table}[h]
\begin{tabular}{cccccccc}\toprule
& &\multicolumn{3}{c}{H}
  &\multicolumn{3}{c}{D}\\\cmidrule(lr){3-5}\cmidrule{6-8}
  TL$_2^{\#}$& $E_B$ & $\Delta_{\rm RPI}^{\rm SP}$& $S/\hbar$ & $\Phi$ &
   $\Delta_{\rm RPI}^{\rm SP}$& $S/\hbar$ & $\Phi$\\\midrule 
0	&	6.619	&	1.01	&	8.68	&	86.36	&	0.055	&	11.23	&	140.87	\\
1	&	6.602	&	1.02	&	8.66	&	87.70	&	0.056	&	11.20	&	144.89	\\
2	&	6.621	&	1.02	&	8.67	&	86.56	&	0.056	&	11.22	&	140.20	\\
3	&	6.625	&	1.03	&	8.68	&	85.09	&	0.056	&	11.23	&	138.18	\\
4	&	6.617	&	1.03	&	8.68	&	85.57	&	0.056	&	11.23	&	139.44	\\
5	&	6.624	&	1.01	&	8.68	&	86.77	&	0.056	&	11.22	&	141.07	\\
6	&	6.626	&	0.97	&	8.68	&	89.95	&	0.053	&	11.24	&	147.37	\\
7	&	6.627	&	0.99	&	8.68	&	88.12	&	0.054	&	11.23	&	144.06	\\
8	&	6.624	&	0.99	&	8.68	&	88.19	&	0.054	&	11.23	&	144.70	\\
9	&	6.619	&	0.98	&	8.68	&	89.43	&	0.054	&	11.23	&	145.52	\\\midrule
$\langle{\rm TL}_2\rangle$	&	6.620	&	1.01	&	8.68	&	87.37	&	0.055	&	11.23	&	142.63	\\
$\sigma$	&	0.007	&	0.02	&	0.01	&	1.59	&	0.001	&	0.01	&	3.05	\\\bottomrule
\end{tabular}
\caption{Energy barriers $E_B$ (kcal/mol), leading order tunneling
  splittings $\Delta_{\rm RPI}^{\rm SP}$ (at $T=25$~K and $N=4096$ beads) given
  in cm$^{-1}$, action $S/\hbar$, fluctuation factor $\Phi$ in a.u. of
  time $\hbar/E_{\rm h}$ and perturbative correction factor $c_{\rm
    PC}$ for tropolone and deuterated tropolone determined from the
    single-precision TL$_2$ PESs.}
\label{sitab:tl2_ensemble_results_tropo_f32}
\end{table}

\begin{table}[h]
\begin{tabular}{cccccccccccc}\toprule
& &\multicolumn{5}{c}{H}
  &\multicolumn{5}{c}{D}\\\cmidrule(lr){3-7}\cmidrule{8-12}
  TL$_2^{\#}$& $E_B$ & $\Delta_{\rm RPI}^{\rm DP}$& $S/\hbar$ & $\Phi$ &
  $c_{\rm PC}$ & $\Delta_{\rm PC}^{\rm DP}$ & $\Delta_{\rm RPI}^{\rm DP}$ & $S/\hbar$ & $\Phi$ & $c_{\rm
    PC}$ &$\Delta_{\rm PC}^{\rm DP}$\\\midrule 
0	&	6.622	&	1.07	&	8.70	&	80.49	&	0.79	&	0.84	&	0.058	&	11.25	&	132.99	&	0.83	&	0.048	\\
1	&	6.609	&	1.10	&	8.67	&	80.69	&	0.86	&	0.95	&	0.059	&	11.21	&	134.31	&	0.82	&	0.048	\\
2	&	6.628	&	1.08	&	8.69	&	80.44	&	0.89	&	0.97	&	0.059	&	11.24	&	131.86	&	0.88	&	0.052	\\
3	&	6.627	&	1.07	&	8.68	&	81.73	&	0.91	&	0.98	&	0.058	&	11.23	&	133.72	&	0.86	&	0.050	\\
4	&	6.622	&	1.05	&	8.68	&	83.07	&	0.86	&	0.91	&	0.057	&	11.23	&	136.91	&	0.57	&	0.032	\\
5	&	6.621	&	1.08	&	8.68	&	81.23	&	0.84	&	0.91	&	0.058	&	11.23	&	134.11	&	0.81	&	0.047	\\
6	&	6.620	&	1.04	&	8.68	&	84.35	&	0.91	&	0.94	&	0.056	&	11.23	&	139.80	&	0.87	&	0.048	\\
7	&	6.629	&	1.05	&	8.68	&	83.55	&	0.96	&	1.01	&	0.057	&	11.23	&	137.06	&	0.90	&	0.051	\\
8	&	6.623	&	1.07	&	8.68	&	82.35	&	0.88	&	0.93	&	0.058	&	11.23	&	135.28	&	0.83	&	0.048	\\
9	&	6.623	&	1.09	&	8.68	&	80.74	&	0.88	&	0.96	&	0.059	&	11.23	&	132.49	&	0.82	&	0.048	\\\midrule
$\langle{\rm TL}_2\rangle$	&	6.622	&	1.07	&	8.68	&	81.86	&	0.88	&	0.94	&	0.058	&	11.23	&	134.85	&	0.82	&	0.047	\\
$\sigma$	&	0.006	&	0.02	&	0.01	&	1.40	&	0.05	&	0.04	&	0.001	&	0.01	&	2.45	&	0.09	&	0.005	\\\bottomrule

\end{tabular}
\caption{Energy barriers $E_B$ (kcal/mol), leading order tunneling
  splittings $\Delta_{\rm RPI}^{\rm DP}$ (at $T=25$~K and $N=4096$ beads) given
  in cm$^{-1}$, action $S/\hbar$, fluctuation factor $\Phi$ in a.u. of
  time $\hbar/E_{\rm h}$ and perturbative correction factor $c_{\rm
    PC}$ for tropolone and deuterated tropolone determined from the
    double-precision TL$_2$ PESs.}
\label{sitab:tl2_ensemble_results_tropo}
\end{table}

\clearpage
\subsection{PFD}

\begin{table}[h]
    \begin{subtable}[c]{0.45\textwidth}
        \centering
\begin{tabular}{rrrr}\toprule
 & \multicolumn{3}{c}{$T = \hbar/(k_B \tau)$}\\\cmidrule(lr){2-4}
 $N$ & 50~K & 25~K & 12.5~K \\\midrule
128	&	0.016	&		&		\\
256	&	0.019	&	0.013	&		\\
512	&	0.020	&	0.015	&	0.013	\\
1024	&	0.020	&	0.016	&	0.015	\\
2048	&	0.020	&	0.016	&	0.016	\\
4096	&	0.020	&	0.016	&	0.016	\\
\bottomrule
\end{tabular}
\caption{Leading order tunneling splitting $\Delta_{\rm RPI}^{\rm DP}$ in
  cm$^{-1}$.}
    \end{subtable}
    \hfill
    \begin{subtable}[c]{0.45\textwidth}
        \centering
\begin{tabular}{rr}\toprule
 & $T = \hbar/(k_B \tau)$\\\cmidrule(lr){2-2}
 $N$ & 50~K  \\\midrule
128	&	0.86  \\
256	&	0.82  \\
512	&	0.82   \\\bottomrule
\end{tabular}
\caption{Perturbative correction factor $c_{\rm PC}$ (unitless).}
     \end{subtable}
\caption{Numerical convergence for double hydrogen transfer in PFD on
  a representative double-precision TL PES}
\label{sitab:convergence_pfd}
\end{table}

\begin{table}[h]
\begin{tabular}{cccccccc}\toprule
& &\multicolumn{3}{c}{H} &\multicolumn{3}{c}{D}\\\cmidrule(lr){3-5}\cmidrule{6-8}
TL$_3^{\#}$& $E_B$ & $\Delta_{\rm RPI}^{\rm SP}$ & $S/\hbar$ & $\Phi$  & $\Delta_{\rm RPI}^{\rm SP}$ & $S/\hbar$ & $\Phi$\\\midrule
0	&	7.762	&	0.0139	&	14.829	&	17.548	&	0.000228	&	18.151	&	42.827	\\
1	&	7.760	&	0.0136	&	14.818	&	18.204	&	0.000226	&	18.134	&	43.852	\\
2	&	7.765	&	0.0154	&	14.837	&	15.732	&	0.000254	&	18.158	&	38.162	\\
3	&	7.762	&	0.0139	&	14.843	&	17.417	&	0.000229	&	18.161	&	42.280	\\
4	&	7.760	&	0.0133	&	14.817	&	18.648	&	0.000222	&	18.132	&	44.876	\\
5	&	7.767	&	0.0150	&	14.825	&	16.411	&	0.000248	&	18.144	&	39.727	\\
6	&	7.758	&	0.0152	&	14.810	&	16.365	&	0.000252	&	18.126	&	39.757	\\
7	&	7.759	&	0.0136	&	14.802	&	18.441	&	0.000228	&	18.117	&	44.267	\\
8	&	7.758	&	0.0140	&	14.822	&	17.629	&	0.000232	&	18.139	&	42.686	\\
9	&	7.754	&	0.0140	&	14.809	&	17.852	&	0.000232	&	18.124	&	43.249	\\\midrule
$\langle {\rm TL}_3\rangle$	&	7.761	&	0.0142	&	14.821	&	17.425	&	0.000235	&	18.139	&	42.168	\\
$\sigma$	&	0.004	&	0.0007	&	0.013	&	0.966	&	0.000012	&	0.015	&	2.218	\\\bottomrule

\end{tabular}
\caption{Energy barriers $E_B$ (kcal/mol), leading order tunneling
  splittings $\Delta_{\rm RPI}^{\rm SP}$ (at $T=25$~K and $N=4096$ beads) given
  in cm$^{-1}$, action $S/\hbar$, fluctuation factor $\Phi$ in a.u. of
  time $\hbar/E_{\rm h}$ and perturbative correction factor $c_{\rm
    PC}$ for PFD and doubly deuterated PFD determined from the 
    single-precision TL$_3$ PESs.}
\label{sitab:tl3_ensemble_results_pfd_f32}
\end{table}

\begin{table}[h]
\begin{tabular}{cccccccccccc}\toprule
& &\multicolumn{5}{c}{H} &\multicolumn{5}{c}{D}\\\cmidrule(lr){3-7}\cmidrule{8-12}
  TL$_3^{\#}$& $E_B$ & $\Delta_{\rm RPI}^{\rm DP}$& $S/\hbar$ & $\Phi$ &
  $c_{\rm PC}$ & $\Delta_{\rm PC}^{\rm DP}$ & $\Delta_{\rm RPI}^{\rm DP}$ & $S/\hbar$ & $\Phi$ & $c_{\rm
    PC}$ &$\Delta_{\rm PC}^{\rm DP}$\\\midrule 
0	&	7.769	&	0.0162	&	14.884	&	14.292	&	0.91	&	0.0148	&	0.000266	&	18.213	&	34.557	&	0.82	&	0.000219	\\
1	&	7.776	&	0.0157	&	14.903	&	14.502	&	0.91	&	0.0142	&	0.000258	&	18.233	&	34.999	&	0.82	&	0.000211	\\
2	&	7.765	&	0.0171	&	14.895	&	13.444	&	0.85	&	0.0145	&	0.000278	&	18.223	&	32.732	&	0.79	&	0.000219	\\
3	&	7.763	&	0.0162	&	14.875	&	14.463	&	0.99	&	0.0160	&	0.000267	&	18.199	&	34.914	&	0.86	&	0.000230	\\
4	&	7.769	&	0.0153	&	14.884	&	15.135	&	0.98	&	0.0150	&	0.000252	&	18.210	&	36.637	&	0.86	&	0.000218	\\
5	&	7.770	&	0.0171	&	14.899	&	13.368	&	0.78	&	0.0133	&	0.000279	&	18.228	&	32.531	&	0.71	&	0.000199	\\
6	&	7.769	&	0.0164	&	14.897	&	13.930	&	0.82	&	0.0135	&	0.000269	&	18.225	&	33.810	&	0.76	&	0.000205	\\
7	&	7.764	&	0.0169	&	14.871	&	13.932	&	0.94	&	0.0159	&	0.000273	&	18.195	&	34.257	&	0.85	&	0.000233	\\
8	&	7.766	&	0.0171	&	14.880	&	13.613	&	0.83	&	0.0143	&	0.000278	&	18.206	&	33.357	&	0.77	&	0.000213	\\
9	&	7.771	&	0.0170	&	14.880	&	13.701	&	0.91	&	0.0154	&	0.000274	&	18.209	&	33.747	&	0.82	&	0.000226	\\\midrule
$\langle {\rm TL}_3\rangle$	&	7.768	&	0.0165	&	14.887	&	14.038	&	0.89	&	0.0147	&	0.000269	&	18.214	&	34.154	&	0.81	&	0.000217	\\
$\sigma$&	0.004	&	0.0006	&	0.011	&	0.556	&	0.07	&	0.0009	&	0.000009	&	0.013	&	1.211	&	0.05	&	0.000011	\\

\bottomrule
\end{tabular}
\caption{Energy barriers $E_B$ (kcal/mol), leading order tunneling
  splittings $\Delta_{\rm RPI}^{\rm DP}$ (at $T=25$~K and $N=4096$ beads) given
  in cm$^{-1}$, action $S/\hbar$, fluctuation factor $\Phi$ in a.u. of
  time $\hbar/E_{\rm h}$ and perturbative correction factor $c_{\rm
    PC}$ for PFD and doubly deuterated PFD determined from the double-precision
    TL$_3$ PESs.}
\label{sitab:tl3_ensemble_results_pfd}
\end{table}
\clearpage

\subsection{FAD}
\begin{table}[h]
    \begin{subtable}[c]{0.45\textwidth}
        \centering
\begin{tabular}{rrrr}\toprule
 & \multicolumn{3}{c}{$T = \hbar/(k_B \tau)$}\\\cmidrule(lr){2-4}
 $N$ & 50~K & 25~K & 12.5~K \\\midrule
128	&	0.0147	&	&	\\		
256	&	0.0175	&	0.0124	&	\\	
512	&	0.0184	&	0.0148	&	0.0124	\\
1024	&	0.0186	&	0.0156	&	0.0148	\\
2048	&	0.0187	&	0.0158	&	0.0155	\\
4096	&	0.0187	&	0.0158	&	0.0157	\\
\bottomrule
\end{tabular}\caption{Leading order tunneling splitting $\Delta_{\rm RPI}^{\rm DP}$
in cm$^{-1}$.}
    \end{subtable}
    \hfill
    \begin{subtable}[c]{0.45\textwidth}
        \centering
\begin{tabular}{rrr}\toprule
 & $T = \hbar/(k_B \tau)$\\\cmidrule(lr){2-2}
 $N$ & 50~K & 25~K \\\midrule
128	&	1.07	&	\\	
256	&	1.17	&	1.05	\\
512	&	1.18	&	1.18	\\\bottomrule
\end{tabular}
\caption{Perturbative correction factor $c_{\rm PC}$ (unitless).}
     \end{subtable}
\caption{Numerical convergence for double hydrogen transfer in FAD on
  the TL (double-precision) PES}
     \label{tab:convergence_fad}
\end{table}

\clearpage
\bibliography{refs}

\providecommand{\latin}[1]{#1}
\makeatletter
\providecommand{\doi}
  {\begingroup\let\do\@makeother\dospecials
  \catcode`\{=1 \catcode`\}=2 \doi@aux}
\providecommand{\doi@aux}[1]{\endgroup\texttt{#1}}
\makeatother
\providecommand*\mcitethebibliography{\thebibliography}
\csname @ifundefined\endcsname{endmcitethebibliography}
  {\let\endmcitethebibliography\endthebibliography}{}
\begin{mcitethebibliography}{89}
\providecommand*\natexlab[1]{#1}
\providecommand*\mciteSetBstSublistMode[1]{}
\providecommand*\mciteSetBstMaxWidthForm[2]{}
\providecommand*\mciteBstWouldAddEndPuncttrue
  {\def\EndOfBibitem{\unskip.}}
\providecommand*\mciteBstWouldAddEndPunctfalse
  {\let\EndOfBibitem\relax}
\providecommand*\mciteSetBstMidEndSepPunct[3]{}
\providecommand*\mciteSetBstSublistLabelBeginEnd[3]{}
\providecommand*\EndOfBibitem{}
\mciteSetBstSublistMode{f}
\mciteSetBstMaxWidthForm{subitem}{(\alph{mcitesubitemcount})}
\mciteSetBstSublistLabelBeginEnd
  {\mcitemaxwidthsubitemform\space}
  {\relax}
  {\relax}

\bibitem[Gol'danskii(2021)]{gol:2021}
Gol'danskii,~V.~I. \emph{Tunneling phenomena in chemical physics}; Routledge,
  2021\relax
\mciteBstWouldAddEndPuncttrue
\mciteSetBstMidEndSepPunct{\mcitedefaultmidpunct}
{\mcitedefaultendpunct}{\mcitedefaultseppunct}\relax
\EndOfBibitem
\bibitem[Rowe~Jr \latin{et~al.}(1976)Rowe~Jr, Duerst, and
  Wilson]{rowe1976intramolecular}
Rowe~Jr,~W.~F.; Duerst,~R.~W.; Wilson,~E.~B. The intramolecular hydrogen bond
  in malonaldehyde. \emph{J. Am. Chem. Soc.} \textbf{1976}, \emph{98},
  4021--4023\relax
\mciteBstWouldAddEndPuncttrue
\mciteSetBstMidEndSepPunct{\mcitedefaultmidpunct}
{\mcitedefaultendpunct}{\mcitedefaultseppunct}\relax
\EndOfBibitem
\bibitem[Ortlieb and Havenith(2007)Ortlieb, and Havenith]{ortlieb2007proton}
Ortlieb,~M.; Havenith,~M. Proton transfer in (HCOOH)$_2$: an IR high-resolution
  spectroscopic study of the antisymmetric C--O stretch. \emph{J. Phys. Chem.
  A} \textbf{2007}, \emph{111}, 7355--7363\relax
\mciteBstWouldAddEndPuncttrue
\mciteSetBstMidEndSepPunct{\mcitedefaultmidpunct}
{\mcitedefaultendpunct}{\mcitedefaultseppunct}\relax
\EndOfBibitem
\bibitem[Taylor and Stone(2009)Taylor, and Stone]{taylor2009transfer}
Taylor,~M.~E.; Stone,~P. Transfer learning for reinforcement learning domains:
  A survey. \emph{J. Mach. Learn. Res.} \textbf{2009}, \emph{10},
  1633--1685\relax
\mciteBstWouldAddEndPuncttrue
\mciteSetBstMidEndSepPunct{\mcitedefaultmidpunct}
{\mcitedefaultendpunct}{\mcitedefaultseppunct}\relax
\EndOfBibitem
\bibitem[Smith \latin{et~al.}(2019)Smith, Nebgen, Zubatyuk, Lubbers, Devereux,
  Barros, Tretiak, Isayev, and Roitberg]{smith2019approaching}
Smith,~J.~S.; Nebgen,~B.~T.; Zubatyuk,~R.; Lubbers,~N.; Devereux,~C.;
  Barros,~K.; Tretiak,~S.; Isayev,~O.; Roitberg,~A.~E. Approaching coupled
  cluster accuracy with a general-purpose neural network potential through
  transfer learning. \emph{Nat. Commun.} \textbf{2019}, \emph{10}, 1--8\relax
\mciteBstWouldAddEndPuncttrue
\mciteSetBstMidEndSepPunct{\mcitedefaultmidpunct}
{\mcitedefaultendpunct}{\mcitedefaultseppunct}\relax
\EndOfBibitem
\bibitem[Pan and Yang(2009)Pan, and Yang]{pan2009survey}
Pan,~S.~J.; Yang,~Q. A survey on transfer learning. \emph{IEEE Trans. Knowl.
  Data Eng.} \textbf{2009}, \emph{22}, 1345--1359\relax
\mciteBstWouldAddEndPuncttrue
\mciteSetBstMidEndSepPunct{\mcitedefaultmidpunct}
{\mcitedefaultendpunct}{\mcitedefaultseppunct}\relax
\EndOfBibitem
\bibitem[Richardson and Althorpe(2011)Richardson, and Althorpe]{tunnel}
Richardson,~J.~O.; Althorpe,~S.~C. Ring-polymer instanton method for
  calculating tunneling splittings. \emph{J.~Chem. Phys.} \textbf{2011},
  \emph{134}, 054109\relax
\mciteBstWouldAddEndPuncttrue
\mciteSetBstMidEndSepPunct{\mcitedefaultmidpunct}
{\mcitedefaultendpunct}{\mcitedefaultseppunct}\relax
\EndOfBibitem
\bibitem[Richardson(2018)]{InstReview}
Richardson,~J.~O. Ring-polymer instanton theory. \emph{Int. Rev. Phys. Chem.}
  \textbf{2018}, \emph{37}, 171--216\relax
\mciteBstWouldAddEndPuncttrue
\mciteSetBstMidEndSepPunct{\mcitedefaultmidpunct}
{\mcitedefaultendpunct}{\mcitedefaultseppunct}\relax
\EndOfBibitem
\bibitem[Richardson \latin{et~al.}(2016)Richardson, P{\'e}rez, Lobsiger, Reid,
  Temelso, Shields, Kisiel, Wales, Pate, and Althorpe]{hexamerprism}
Richardson,~J.~O.; P{\'e}rez,~C.; Lobsiger,~S.; Reid,~A.~A.; Temelso,~B.;
  Shields,~G.~C.; Kisiel,~Z.; Wales,~D.~J.; Pate,~B.~H.; Althorpe,~S.~C.
  Concerted Hydrogen-Bond Breaking by Quantum Tunneling in the Water Hexamer
  Prism. \emph{Science} \textbf{2016}, \emph{351}, 1310--1313\relax
\mciteBstWouldAddEndPuncttrue
\mciteSetBstMidEndSepPunct{\mcitedefaultmidpunct}
{\mcitedefaultendpunct}{\mcitedefaultseppunct}\relax
\EndOfBibitem
\bibitem[Richardson(2018)]{Perspective}
Richardson,~J.~O. {Perspective: Ring-polymer instanton theory}. \emph{J. Chem.
  Phys.} \textbf{2018}, \emph{148}, 200901\relax
\mciteBstWouldAddEndPuncttrue
\mciteSetBstMidEndSepPunct{\mcitedefaultmidpunct}
{\mcitedefaultendpunct}{\mcitedefaultseppunct}\relax
\EndOfBibitem
\bibitem[Lawrence \latin{et~al.}(2023)Lawrence, Dušek, and
  Richardson]{richardson2023pcrpi}
Lawrence,~J.~E.; Dušek,~J.; Richardson,~J.~O. {Perturbatively corrected
  ring-polymer instanton theory for accurate tunneling splittings}. \emph{J.
  Chem. Phys.} \textbf{2023}, \emph{159}, 014111\relax
\mciteBstWouldAddEndPuncttrue
\mciteSetBstMidEndSepPunct{\mcitedefaultmidpunct}
{\mcitedefaultendpunct}{\mcitedefaultseppunct}\relax
\EndOfBibitem
\bibitem[K{\"a}ser \latin{et~al.}(2022)K{\"a}ser, Richardson, and
  Meuwly]{mm.tlrpimalonaldehyde:2022}
K{\"a}ser,~S.; Richardson,~J.~O.; Meuwly,~M. Transfer Learning for Affordable
  and High-Quality Tunneling Splittings from Instanton Calculations. \emph{J.
  Chem. Theory Comput.} \textbf{2022}, \emph{18}, 6840--6850\relax
\mciteBstWouldAddEndPuncttrue
\mciteSetBstMidEndSepPunct{\mcitedefaultmidpunct}
{\mcitedefaultendpunct}{\mcitedefaultseppunct}\relax
\EndOfBibitem
\bibitem[Firth \latin{et~al.}(1991)Firth, Beyer, Dvorak, Reeve, Grushow, and
  Leopold]{firth1991tunable}
Firth,~D.; Beyer,~K.; Dvorak,~M.; Reeve,~S.; Grushow,~A.; Leopold,~K. Tunable
  far-infrared spectroscopy of malonaldehyde. \emph{J. Chem. Phys.}
  \textbf{1991}, \emph{94}, 1812--1819\relax
\mciteBstWouldAddEndPuncttrue
\mciteSetBstMidEndSepPunct{\mcitedefaultmidpunct}
{\mcitedefaultendpunct}{\mcitedefaultseppunct}\relax
\EndOfBibitem
\bibitem[Baba \latin{et~al.}(1999)Baba, Tanaka, Morino, Yamada, and
  Tanaka]{baba1999detection}
Baba,~T.; Tanaka,~T.; Morino,~I.; Yamada,~K.~M.; Tanaka,~K. Detection of the
  tunneling-rotation transitions of malonaldehyde in the submillimeter-wave
  region. \emph{J. Chem. Phys.} \textbf{1999}, \emph{110}, 4131--4133\relax
\mciteBstWouldAddEndPuncttrue
\mciteSetBstMidEndSepPunct{\mcitedefaultmidpunct}
{\mcitedefaultendpunct}{\mcitedefaultseppunct}\relax
\EndOfBibitem
\bibitem[Baughcum \latin{et~al.}(1984)Baughcum, Smith, Wilson, and
  Duerst]{baughcum1984microwave}
Baughcum,~S.~L.; Smith,~Z.; Wilson,~E.~B.; Duerst,~R.~W. Microwave
  spectroscopic study of malonaldehyde. 3. Vibration-rotation interaction and
  one-dimensional model for proton tunneling. \emph{J. Am. Chem. Soc.}
  \textbf{1984}, \emph{106}, 2260--2265\relax
\mciteBstWouldAddEndPuncttrue
\mciteSetBstMidEndSepPunct{\mcitedefaultmidpunct}
{\mcitedefaultendpunct}{\mcitedefaultseppunct}\relax
\EndOfBibitem
\bibitem[Turner \latin{et~al.}(1984)Turner, Baughcum, Coy, and
  Smith]{turner1984microwave}
Turner,~P.; Baughcum,~S.~L.; Coy,~S.~L.; Smith,~Z. Microwave spectroscopic
  study of malonaldehyde. 4. Vibration-rotation interaction in parent species.
  \emph{J. Am. Chem. Soc.} \textbf{1984}, \emph{106}, 2265--2267\relax
\mciteBstWouldAddEndPuncttrue
\mciteSetBstMidEndSepPunct{\mcitedefaultmidpunct}
{\mcitedefaultendpunct}{\mcitedefaultseppunct}\relax
\EndOfBibitem
\bibitem[Smith \latin{et~al.}(1983)Smith, Wilson, and
  Duerst]{smith1983infrared}
Smith,~Z.; Wilson,~E.~B.; Duerst,~R.~W. The infrared spectrum of gaseous
  malonaldehyde (3-hydroxy-2-propenal). \emph{Spectrochim. Acta A Mol. Biomol.
  Spectrosc.} \textbf{1983}, \emph{39}, 1117--1129\relax
\mciteBstWouldAddEndPuncttrue
\mciteSetBstMidEndSepPunct{\mcitedefaultmidpunct}
{\mcitedefaultendpunct}{\mcitedefaultseppunct}\relax
\EndOfBibitem
\bibitem[Firth \latin{et~al.}(1989)Firth, Barbara, and
  Trommsdorff]{firth1989matrix}
Firth,~D.~W.; Barbara,~P.~F.; Trommsdorff,~H.~P. Matrix induced localization of
  proton tunneling in malonaldehyde. \emph{Chem. Phys.} \textbf{1989},
  \emph{136}, 349--360\relax
\mciteBstWouldAddEndPuncttrue
\mciteSetBstMidEndSepPunct{\mcitedefaultmidpunct}
{\mcitedefaultendpunct}{\mcitedefaultseppunct}\relax
\EndOfBibitem
\bibitem[Chiavassa \latin{et~al.}(1992)Chiavassa, Roubin, Pizzala, Verlaque,
  Allouche, and Marinelli]{chiavassa1992experimental}
Chiavassa,~T.; Roubin,~P.; Pizzala,~L.; Verlaque,~P.; Allouche,~A.;
  Marinelli,~F. Experimental and theoretical studies of malonaldehyde:
  Vibrational analysis of a strongly intramolecularly hydrogen bonded compound.
  \emph{J. Chem. Phys.} \textbf{1992}, \emph{96}, 10659--10665\relax
\mciteBstWouldAddEndPuncttrue
\mciteSetBstMidEndSepPunct{\mcitedefaultmidpunct}
{\mcitedefaultendpunct}{\mcitedefaultseppunct}\relax
\EndOfBibitem
\bibitem[Duan and Luckhaus(2004)Duan, and Luckhaus]{duan2004high}
Duan,~C.; Luckhaus,~D. High resolution IR-diode laser jet spectroscopy of
  malonaldehyde. \emph{Chem. Phys. Lett.} \textbf{2004}, \emph{391},
  129--133\relax
\mciteBstWouldAddEndPuncttrue
\mciteSetBstMidEndSepPunct{\mcitedefaultmidpunct}
{\mcitedefaultendpunct}{\mcitedefaultseppunct}\relax
\EndOfBibitem
\bibitem[Wang \latin{et~al.}(2008)Wang, Braams, Bowman, Carter, and
  Tew]{wang2008full}
Wang,~Y.; Braams,~B.~J.; Bowman,~J.~M.; Carter,~S.; Tew,~D.~P. Full-dimensional
  quantum calculations of ground-state tunneling splitting of malonaldehyde
  using an accurate ab initio potential energy surface. \emph{J. Chem. Phys.}
  \textbf{2008}, \emph{128}, 224314\relax
\mciteBstWouldAddEndPuncttrue
\mciteSetBstMidEndSepPunct{\mcitedefaultmidpunct}
{\mcitedefaultendpunct}{\mcitedefaultseppunct}\relax
\EndOfBibitem
\bibitem[Yang and Meuwly(2010)Yang, and Meuwly]{MM.ma:2010}
Yang,~Y.; Meuwly,~M. A generalized reactive force field for nonlinear hydrogen
  bonds: Hydrogen dynamics and transfer in malonaldehyde. \emph{J. Chem. Phys.}
  \textbf{2010}, \emph{133}, 064503\relax
\mciteBstWouldAddEndPuncttrue
\mciteSetBstMidEndSepPunct{\mcitedefaultmidpunct}
{\mcitedefaultendpunct}{\mcitedefaultseppunct}\relax
\EndOfBibitem
\bibitem[Mizukami \latin{et~al.}(2014)Mizukami, Habershon, and
  Tew]{mizukami2014compact}
Mizukami,~W.; Habershon,~S.; Tew,~D.~P. A compact and accurate semi-global
  potential energy surface for malonaldehyde from constrained least squares
  regression. \emph{J. Chem. Phys.} \textbf{2014}, \emph{141}, 144310\relax
\mciteBstWouldAddEndPuncttrue
\mciteSetBstMidEndSepPunct{\mcitedefaultmidpunct}
{\mcitedefaultendpunct}{\mcitedefaultseppunct}\relax
\EndOfBibitem
\bibitem[Huang \latin{et~al.}(2014)Huang, Buchowiecki, Nagy,
  Van{\'\i}{\v{c}}ek, and Meuwly]{MM.ma:2014}
Huang,~J.; Buchowiecki,~M.; Nagy,~T.; Van{\'\i}{\v{c}}ek,~J.; Meuwly,~M.
  Kinetic isotope effect in malonaldehyde determined from path integral Monte
  Carlo simulations. \emph{Phys. Chem. Chem. Phys.} \textbf{2014}, \emph{16},
  204--211\relax
\mciteBstWouldAddEndPuncttrue
\mciteSetBstMidEndSepPunct{\mcitedefaultmidpunct}
{\mcitedefaultendpunct}{\mcitedefaultseppunct}\relax
\EndOfBibitem
\bibitem[Cvitas and Althorpe(2016)Cvitas, and Althorpe]{cvitas2016locating}
Cvitas,~M.~T.; Althorpe,~S.~C. Locating instantons in calculations of tunneling
  splittings: The test case of malonaldehyde. \emph{J. Chem. Theory Comput.}
  \textbf{2016}, \emph{12}, 787--803\relax
\mciteBstWouldAddEndPuncttrue
\mciteSetBstMidEndSepPunct{\mcitedefaultmidpunct}
{\mcitedefaultendpunct}{\mcitedefaultseppunct}\relax
\EndOfBibitem
\bibitem[K{\"a}ser \latin{et~al.}(2020)K{\"a}ser, Unke, and Meuwly]{mm.ht:2020}
K{\"a}ser,~S.; Unke,~O.~T.; Meuwly,~M. Reactive dynamics and spectroscopy of
  hydrogen transfer from neural network-based reactive potential energy
  surfaces. \emph{New J. Phys.} \textbf{2020}, \emph{22}, 055002\relax
\mciteBstWouldAddEndPuncttrue
\mciteSetBstMidEndSepPunct{\mcitedefaultmidpunct}
{\mcitedefaultendpunct}{\mcitedefaultseppunct}\relax
\EndOfBibitem
\bibitem[Jahr \latin{et~al.}(2020)Jahr, Laude, and
  Richardson]{jahr2020instanton}
Jahr,~E.; Laude,~G.; Richardson,~J.~O. Instanton theory of tunneling in
  molecules with asymmetric isotopic substitutions. \emph{J. Chem. Phys.}
  \textbf{2020}, \emph{153}, 094101\relax
\mciteBstWouldAddEndPuncttrue
\mciteSetBstMidEndSepPunct{\mcitedefaultmidpunct}
{\mcitedefaultendpunct}{\mcitedefaultseppunct}\relax
\EndOfBibitem
\bibitem[Ito and Nakanaga(2000)Ito, and Nakanaga]{ito2000jet}
Ito,~F.; Nakanaga,~T. A jet-cooled infrared spectrum of the formic acid dimer
  by cavity ring-down spectroscopy. \emph{Chem. Phys. Lett.} \textbf{2000},
  \emph{318}, 571--577\relax
\mciteBstWouldAddEndPuncttrue
\mciteSetBstMidEndSepPunct{\mcitedefaultmidpunct}
{\mcitedefaultendpunct}{\mcitedefaultseppunct}\relax
\EndOfBibitem
\bibitem[Georges \latin{et~al.}(2004)Georges, Freytes, Hurtmans, Kleiner,
  Vander~Auwera, and Herman]{georges2004jet}
Georges,~R.; Freytes,~M.; Hurtmans,~D.; Kleiner,~I.; Vander~Auwera,~J.;
  Herman,~M. Jet-cooled and room temperature FTIR spectra of the dimer of
  formic acid in the gas phase. \emph{Chem. Phys.} \textbf{2004}, \emph{305},
  187--196\relax
\mciteBstWouldAddEndPuncttrue
\mciteSetBstMidEndSepPunct{\mcitedefaultmidpunct}
{\mcitedefaultendpunct}{\mcitedefaultseppunct}\relax
\EndOfBibitem
\bibitem[Zielke and Suhm(2007)Zielke, and Suhm]{zielke:2007}
Zielke,~P.; Suhm,~M. Raman jet spectroscopy of formic acid dimers: low
  frequency vibrational dynamics and beyond. \emph{Phys. Chem. Chem. Phys.}
  \textbf{2007}, \emph{9}, 4528--4534\relax
\mciteBstWouldAddEndPuncttrue
\mciteSetBstMidEndSepPunct{\mcitedefaultmidpunct}
{\mcitedefaultendpunct}{\mcitedefaultseppunct}\relax
\EndOfBibitem
\bibitem[Xue and Suhm(2009)Xue, and Suhm]{xue2009probing}
Xue,~Z.; Suhm,~M. Probing the stiffness of the simplest double hydrogen bond:
  The symmetric hydrogen bond modes of jet-cooled formic acid dimer. \emph{J.
  Chem. Phys.} \textbf{2009}, \emph{131}, 054301\relax
\mciteBstWouldAddEndPuncttrue
\mciteSetBstMidEndSepPunct{\mcitedefaultmidpunct}
{\mcitedefaultendpunct}{\mcitedefaultseppunct}\relax
\EndOfBibitem
\bibitem[Kollipost \latin{et~al.}(2012)Kollipost, Larsen, Domanskaya,
  Nörenberg, and Suhm]{suhm:2012}
Kollipost,~F.; Larsen,~R.~W.; Domanskaya,~A.~V.; Nörenberg,~M.; Suhm,~M.~A.
  Communication: The highest frequency hydrogen bond vibration and an
  experimental value for the dissociation energy of formic acid dimer. \emph{J.
  Chem. Phys.} \textbf{2012}, \emph{136}, 151101\relax
\mciteBstWouldAddEndPuncttrue
\mciteSetBstMidEndSepPunct{\mcitedefaultmidpunct}
{\mcitedefaultendpunct}{\mcitedefaultseppunct}\relax
\EndOfBibitem
\bibitem[Goroya \latin{et~al.}(2014)Goroya, Zhu, Sun, and Duan]{goroya2014high}
Goroya,~K.~G.; Zhu,~Y.; Sun,~P.; Duan,~C. High resolution jet-cooled infrared
  absorption spectra of the formic acid dimer: a reinvestigation of the C--O
  stretch region. \emph{J. Chem. Phys.} \textbf{2014}, \emph{140}, 164311\relax
\mciteBstWouldAddEndPuncttrue
\mciteSetBstMidEndSepPunct{\mcitedefaultmidpunct}
{\mcitedefaultendpunct}{\mcitedefaultseppunct}\relax
\EndOfBibitem
\bibitem[Zhang \latin{et~al.}(2017)Zhang, Li, Luo, Zhu, and Duan]{duan:2017}
Zhang,~Y.; Li,~W.; Luo,~W.; Zhu,~Y.; Duan,~C. High resolution jet-cooled
  infrared absorption spectra of (HCOOH)$_2$,(HCOOD)$_2$, and HCOOH-HCOOD
  complexes in 7.2 $\mu$m region. \emph{J. Chem. Phys.} \textbf{2017},
  \emph{146}, 244306\relax
\mciteBstWouldAddEndPuncttrue
\mciteSetBstMidEndSepPunct{\mcitedefaultmidpunct}
{\mcitedefaultendpunct}{\mcitedefaultseppunct}\relax
\EndOfBibitem
\bibitem[Li \latin{et~al.}(2019)Li, Evangelisti, Gou, Caminati, and
  Meyer]{li2019barrier}
Li,~W.; Evangelisti,~L.; Gou,~Q.; Caminati,~W.; Meyer,~R. The barrier to proton
  transfer in the dimer of formic acid: A pure rotational study. \emph{Angew.
  Chem. Int. Ed.} \textbf{2019}, \emph{58}, 859--865\relax
\mciteBstWouldAddEndPuncttrue
\mciteSetBstMidEndSepPunct{\mcitedefaultmidpunct}
{\mcitedefaultendpunct}{\mcitedefaultseppunct}\relax
\EndOfBibitem
\bibitem[Nejad and Suhm(2020)Nejad, and Suhm]{suhm:2020}
Nejad,~A.; Suhm,~M.~A. Concerted pair motion due to double hydrogen bonding:
  The formic acid dimer case. \emph{J. Ind. Inst. Sci.} \textbf{2020},
  \emph{100}, 5--19\relax
\mciteBstWouldAddEndPuncttrue
\mciteSetBstMidEndSepPunct{\mcitedefaultmidpunct}
{\mcitedefaultendpunct}{\mcitedefaultseppunct}\relax
\EndOfBibitem
\bibitem[Kalescky \latin{et~al.}(2013)Kalescky, Kraka, and
  Cremer]{kalescky2013local}
Kalescky,~R.; Kraka,~E.; Cremer,~D. Local vibrational modes of the formic acid
  dimer--the strength of the double hydrogen bond. \emph{Mol. Phys.}
  \textbf{2013}, \emph{111}, 1497--1510\relax
\mciteBstWouldAddEndPuncttrue
\mciteSetBstMidEndSepPunct{\mcitedefaultmidpunct}
{\mcitedefaultendpunct}{\mcitedefaultseppunct}\relax
\EndOfBibitem
\bibitem[Ivanov \latin{et~al.}(2015)Ivanov, Grant, and Marx]{ivanov2015quantum}
Ivanov,~S.~D.; Grant,~I.~M.; Marx,~D. Quantum free energy landscapes from ab
  initio path integral metadynamics: Double proton transfer in the formic acid
  dimer is concerted but not correlated. \emph{J. Chem. Phys.} \textbf{2015},
  \emph{143}, 124304\relax
\mciteBstWouldAddEndPuncttrue
\mciteSetBstMidEndSepPunct{\mcitedefaultmidpunct}
{\mcitedefaultendpunct}{\mcitedefaultseppunct}\relax
\EndOfBibitem
\bibitem[Miliordos and Xantheas(2015)Miliordos, and
  Xantheas]{miliordos2015validity}
Miliordos,~E.; Xantheas,~S.~S. On the validity of the basis set superposition
  error and complete basis set limit extrapolations for the binding energy of
  the formic acid dimer. \emph{J. Chem. Phys.} \textbf{2015}, \emph{142},
  094311\relax
\mciteBstWouldAddEndPuncttrue
\mciteSetBstMidEndSepPunct{\mcitedefaultmidpunct}
{\mcitedefaultendpunct}{\mcitedefaultseppunct}\relax
\EndOfBibitem
\bibitem[Tew and Mizukami(2016)Tew, and Mizukami]{tew2016ab}
Tew,~D.~P.; Mizukami,~W. Ab initio vibrational spectroscopy of cis-and
  trans-formic acid from a global potential energy surface. \emph{J. Phys.
  Chem. A} \textbf{2016}, \emph{120}, 9815--9828\relax
\mciteBstWouldAddEndPuncttrue
\mciteSetBstMidEndSepPunct{\mcitedefaultmidpunct}
{\mcitedefaultendpunct}{\mcitedefaultseppunct}\relax
\EndOfBibitem
\bibitem[Qu and Bowman(2016)Qu, and Bowman]{qu2016ab}
Qu,~C.; Bowman,~J.~M. An ab initio potential energy surface for the formic acid
  dimer: zero-point energy, selected anharmonic fundamental energies, and
  ground-state tunneling splitting calculated in relaxed 1--4-mode subspaces.
  \emph{Phys. Chem. Chem. Phys.} \textbf{2016}, \emph{18}, 24835--24840\relax
\mciteBstWouldAddEndPuncttrue
\mciteSetBstMidEndSepPunct{\mcitedefaultmidpunct}
{\mcitedefaultendpunct}{\mcitedefaultseppunct}\relax
\EndOfBibitem
\bibitem[Mackeprang \latin{et~al.}(2016)Mackeprang, Xu, Maroun, Meuwly, and
  Kjaergaard]{MM.fad:2016}
Mackeprang,~K.; Xu,~Z.-H.; Maroun,~Z.; Meuwly,~M.; Kjaergaard,~H.~G.
  Spectroscopy and dynamics of double proton transfer in formic acid dimer.
  \emph{Phys. Chem. Chem. Phys.} \textbf{2016}, \emph{18}, 24654--24662\relax
\mciteBstWouldAddEndPuncttrue
\mciteSetBstMidEndSepPunct{\mcitedefaultmidpunct}
{\mcitedefaultendpunct}{\mcitedefaultseppunct}\relax
\EndOfBibitem
\bibitem[Richardson(2017)]{richardson2017full}
Richardson,~J.~O. Full-and reduced-dimensionality instanton calculations of the
  tunnelling splitting in the formic acid dimer. \emph{Phys. Chem. Chem. Phys.}
  \textbf{2017}, \emph{19}, 966--970\relax
\mciteBstWouldAddEndPuncttrue
\mciteSetBstMidEndSepPunct{\mcitedefaultmidpunct}
{\mcitedefaultendpunct}{\mcitedefaultseppunct}\relax
\EndOfBibitem
\bibitem[Qu and Bowman(2018)Qu, and Bowman]{qu2018high}
Qu,~C.; Bowman,~J.~M. High-dimensional fitting of sparse datasets of CCSD(T)
  electronic energies and MP2 dipole moments, illustrated for the formic acid
  dimer and its complex IR spectrum. \emph{J. Chem. Phys.} \textbf{2018},
  \emph{148}, 241713\relax
\mciteBstWouldAddEndPuncttrue
\mciteSetBstMidEndSepPunct{\mcitedefaultmidpunct}
{\mcitedefaultendpunct}{\mcitedefaultseppunct}\relax
\EndOfBibitem
\bibitem[Qu and Bowman(2018)Qu, and Bowman]{qu2018quantum}
Qu,~C.; Bowman,~J.~M. Quantum and classical IR spectra of
  (HCOOH)$_2$,(DCOOH)$_2$ and (DCOOD)$_2$ using ab initio potential energy and
  dipole moment surfaces. \emph{Faraday Discuss.} \textbf{2018}, \emph{212},
  33--49\relax
\mciteBstWouldAddEndPuncttrue
\mciteSetBstMidEndSepPunct{\mcitedefaultmidpunct}
{\mcitedefaultendpunct}{\mcitedefaultseppunct}\relax
\EndOfBibitem
\bibitem[Qu and Bowman(2018)Qu, and Bowman]{qu2018ir}
Qu,~C.; Bowman,~J.~M. IR Spectra of (HCOOH)$_2$ and (DCOOH)$_2$: Experiment,
  VSCF/VCI, and ab initio molecular dynamics calculations using
  full-dimensional potential and dipole moment surfaces. \emph{J. Phys. Chem.
  Lett} \textbf{2018}, \emph{9}, 2604--2610\relax
\mciteBstWouldAddEndPuncttrue
\mciteSetBstMidEndSepPunct{\mcitedefaultmidpunct}
{\mcitedefaultendpunct}{\mcitedefaultseppunct}\relax
\EndOfBibitem
\bibitem[K\"aser and Meuwly(2023)K\"aser, and Meuwly]{kaeser2023numerical}
K\"aser,~S.; Meuwly,~M. Numerical Accuracy Matters: Applications of Machine
  Learned Potential Energy Surfaces. \emph{J. Phys. Chem. Lett.} \textbf{2023},
  \emph{15}, 3419--3424\relax
\mciteBstWouldAddEndPuncttrue
\mciteSetBstMidEndSepPunct{\mcitedefaultmidpunct}
{\mcitedefaultendpunct}{\mcitedefaultseppunct}\relax
\EndOfBibitem
\bibitem[Tanaka \latin{et~al.}(1999)Tanaka, Honjo, Tanaka, Kohguchi, Ohshima,
  and Endo]{tanaka1999determination}
Tanaka,~K.; Honjo,~H.; Tanaka,~T.; Kohguchi,~H.; Ohshima,~Y.; Endo,~Y.
  Determination of the proton tunneling splitting of tropolone in the ground
  state by microwave spectroscopy. \emph{J. Chem. Phys.} \textbf{1999},
  \emph{110}, 1969--1978\relax
\mciteBstWouldAddEndPuncttrue
\mciteSetBstMidEndSepPunct{\mcitedefaultmidpunct}
{\mcitedefaultendpunct}{\mcitedefaultseppunct}\relax
\EndOfBibitem
\bibitem[Keske \latin{et~al.}(2006)Keske, Lin, Pringle, Novick, Blake, and
  Plusquellic]{keske2006highresolution}
Keske,~J.~C.; Lin,~W.; Pringle,~W.~C.; Novick,~S.~E.; Blake,~T.~A.;
  Plusquellic,~D.~F. High-resolution studies of tropolone in the {$S_0$} and
  {$S_1$} electronic states: {Isotope} driven dynamics in the zero-point energy
  levels. \emph{J. Chem. Phys.} \textbf{2006}, \emph{124}, 074309\relax
\mciteBstWouldAddEndPuncttrue
\mciteSetBstMidEndSepPunct{\mcitedefaultmidpunct}
{\mcitedefaultendpunct}{\mcitedefaultseppunct}\relax
\EndOfBibitem
\bibitem[Houston \latin{et~al.}(2020)Houston, Conte, Qu, and
  Bowman]{houston2020permutationally}
Houston,~P.; Conte,~R.; Qu,~C.; Bowman,~J.~M. Permutationally invariant
  polynomial potential energy surfaces for tropolone and H and D atom tunneling
  dynamics. \emph{J. Chem. Phys.} \textbf{2020}, \emph{153}, 024107\relax
\mciteBstWouldAddEndPuncttrue
\mciteSetBstMidEndSepPunct{\mcitedefaultmidpunct}
{\mcitedefaultendpunct}{\mcitedefaultseppunct}\relax
\EndOfBibitem
\bibitem[Sahu and Gadre(2014)Sahu, and Gadre]{sahu2014molecular}
Sahu,~N.; Gadre,~S.~R. Molecular tailoring approach: a route for ab initio
  treatment of large clusters. \emph{Acc. Chem. Res.} \textbf{2014}, \emph{47},
  2739--2747\relax
\mciteBstWouldAddEndPuncttrue
\mciteSetBstMidEndSepPunct{\mcitedefaultmidpunct}
{\mcitedefaultendpunct}{\mcitedefaultseppunct}\relax
\EndOfBibitem
\bibitem[Nandi \latin{et~al.}(2023)Nandi, Laude, Khire, Gurav, Qu, Conte, Yu,
  Li, Houston, Gadre, \latin{et~al.} others]{nandi2023ring}
Nandi,~A.; Laude,~G.; Khire,~S.~S.; Gurav,~N.~D.; Qu,~C.; Conte,~R.; Yu,~Q.;
  Li,~S.; Houston,~P.~L.; Gadre,~S.~R. \latin{et~al.}  Ring-Polymer Instanton
  Tunneling Splittings of Tropolone and Isotopomers using a $\Delta$-Machine
  Learned CCSD(T) Potential: Theory and Experiment Shake Hands. \emph{J. Am.
  Chem. Soc.} \textbf{2023}, \emph{145}, 9655--9664\relax
\mciteBstWouldAddEndPuncttrue
\mciteSetBstMidEndSepPunct{\mcitedefaultmidpunct}
{\mcitedefaultendpunct}{\mcitedefaultseppunct}\relax
\EndOfBibitem
\bibitem[Daly \latin{et~al.}(2011)Daly, Douglass, Sarkozy, Neill, Muckle,
  Zaleski, Pate, and Kukolich]{daly2011microwave}
Daly,~A.~M.; Douglass,~K.~O.; Sarkozy,~L.~C.; Neill,~J.~L.; Muckle,~M.~T.;
  Zaleski,~D.~P.; Pate,~B.~H.; Kukolich,~S.~G. Microwave measurements of proton
  tunneling and structural parameters for the propiolic acid--formic acid
  dimer. \emph{J. Chem. Phys.} \textbf{2011}, \emph{135}, 154304\relax
\mciteBstWouldAddEndPuncttrue
\mciteSetBstMidEndSepPunct{\mcitedefaultmidpunct}
{\mcitedefaultendpunct}{\mcitedefaultseppunct}\relax
\EndOfBibitem
\bibitem[Meyer and Nejad(2021)Meyer, and Nejad]{meyer2021cc}
Meyer,~K.~A.; Nejad,~A. CC-stretched formic acid: isomerisation, dimerisation,
  and carboxylic acid complexation. \emph{Phys. Chem. Chem. Phys.}
  \textbf{2021}, \emph{23}, 17208--17223\relax
\mciteBstWouldAddEndPuncttrue
\mciteSetBstMidEndSepPunct{\mcitedefaultmidpunct}
{\mcitedefaultendpunct}{\mcitedefaultseppunct}\relax
\EndOfBibitem
\bibitem[Sun \latin{et~al.}(2013)Sun, Wang, Carey, Mitchell, Bowman, and
  Kukolich]{sun2013calculations}
Sun,~M.; Wang,~Y.; Carey,~S.~J.; Mitchell,~E.~G.; Bowman,~J.; Kukolich,~S.~G.
  Calculations and measurements of the deuterium tunneling frequency in the
  propiolic acid-formic acid dimer and description of a newly constructed
  Fourier transform microwave spectrometer. \emph{J. Chem. Phys.}
  \textbf{2013}, \emph{139}, 084316\relax
\mciteBstWouldAddEndPuncttrue
\mciteSetBstMidEndSepPunct{\mcitedefaultmidpunct}
{\mcitedefaultendpunct}{\mcitedefaultseppunct}\relax
\EndOfBibitem
\bibitem[Tanaka \latin{et~al.}(1999)Tanaka, Honjo, Tanaka, Kohguchi, Ohshima,
  and Endo]{tanaka:1999}
Tanaka,~K.; Honjo,~H.; Tanaka,~T.; Kohguchi,~H.; Ohshima,~Y.; Endo,~Y.
  Determination of the proton tunneling splitting of tropolone in the ground
  state by microwave spectroscopy. \emph{J. Chem. Phys.} \textbf{1999},
  \emph{110}, 1969--1978\relax
\mciteBstWouldAddEndPuncttrue
\mciteSetBstMidEndSepPunct{\mcitedefaultmidpunct}
{\mcitedefaultendpunct}{\mcitedefaultseppunct}\relax
\EndOfBibitem
\bibitem[K{\"a}ser \latin{et~al.}(2021)K{\"a}ser, Boittier, Upadhyay, and
  Meuwly]{mm.anharmonic:2021}
K{\"a}ser,~S.; Boittier,~E.~D.; Upadhyay,~M.; Meuwly,~M. Transfer Learning to
  CCSD(T): Accurate Anharmonic Frequencies from Machine Learning Models.
  \emph{J. Chem. Theory Comput.} \textbf{2021}, \emph{17}, 3687--3699\relax
\mciteBstWouldAddEndPuncttrue
\mciteSetBstMidEndSepPunct{\mcitedefaultmidpunct}
{\mcitedefaultendpunct}{\mcitedefaultseppunct}\relax
\EndOfBibitem
\bibitem[Khire \latin{et~al.}(2022)Khire, Gurav, Nandi, and
  Gadre]{khire2022enabling}
Khire,~S.~S.; Gurav,~N.~D.; Nandi,~A.; Gadre,~S.~R. Enabling rapid and accurate
  construction of CCSD (T)-level potential energy surface of large molecules
  using molecular tailoring approach. \emph{J. Phys. Chem. A} \textbf{2022},
  \emph{126}, 1458--1464\relax
\mciteBstWouldAddEndPuncttrue
\mciteSetBstMidEndSepPunct{\mcitedefaultmidpunct}
{\mcitedefaultendpunct}{\mcitedefaultseppunct}\relax
\EndOfBibitem
\bibitem[M{\'a}tyus \latin{et~al.}(2016)M{\'a}tyus, Wales, and
  Althorpe]{Matyus2016tunnel1}
M{\'a}tyus,~E.; Wales,~D.~J.; Althorpe,~S.~C. Quantum tunneling splittings from
  path-integral molecular dynamics. \emph{J.~Chem. Phys.} \textbf{2016},
  \emph{144}, 114108\relax
\mciteBstWouldAddEndPuncttrue
\mciteSetBstMidEndSepPunct{\mcitedefaultmidpunct}
{\mcitedefaultendpunct}{\mcitedefaultseppunct}\relax
\EndOfBibitem
\bibitem[Vaillant \latin{et~al.}(2018)Vaillant, Wales, and
  Althorpe]{Vaillant2018dimer}
Vaillant,~C.~L.; Wales,~D.~J.; Althorpe,~S.~C. Tunneling splittings from
  path-integral molecular dynamics using a {L}angevin thermostat.
  \emph{J.~Chem. Phys.} \textbf{2018}, \emph{148}, 234102\relax
\mciteBstWouldAddEndPuncttrue
\mciteSetBstMidEndSepPunct{\mcitedefaultmidpunct}
{\mcitedefaultendpunct}{\mcitedefaultseppunct}\relax
\EndOfBibitem
\bibitem[Trenins \latin{et~al.}(2023)Trenins, Meuser, Bertschi, Vavourakis,
  Fl\"utsch, and Richardson]{PIMDtunnel}
Trenins,~G.; Meuser,~L.; Bertschi,~H.; Vavourakis,~O.; Fl\"utsch,~R.;
  Richardson,~J.~O. Exact tunnelling splittings from symmetrized path
  integrals. \emph{J. Chem. Phys.} \textbf{2023}, \emph{159}, 034108\relax
\mciteBstWouldAddEndPuncttrue
\mciteSetBstMidEndSepPunct{\mcitedefaultmidpunct}
{\mcitedefaultendpunct}{\mcitedefaultseppunct}\relax
\EndOfBibitem
\bibitem[Kosztin \latin{et~al.}(1996)Kosztin, Faber, and
  Schulten]{kosztin1996introduction}
Kosztin,~I.; Faber,~B.; Schulten,~K. Introduction to the diffusion Monte Carlo
  method. \emph{Am. J. Phys.} \textbf{1996}, \emph{64}, 633--644\relax
\mciteBstWouldAddEndPuncttrue
\mciteSetBstMidEndSepPunct{\mcitedefaultmidpunct}
{\mcitedefaultendpunct}{\mcitedefaultseppunct}\relax
\EndOfBibitem
\bibitem[Redington \latin{et~al.}(2008)Redington, Redington, and
  Sams]{redington2008tunnelingtropo}
Redington,~R.~L.; Redington,~T.~E.; Sams,~R.~L. Tunneling {Splittings} for
  “{O}···{O} {Stretching}” and {Other} {Vibrations} of {Tropolone}
  {Isotopomers} {Observed} in the {Infrared} {Spectrum} {Below} 800 cm
  $^{\textrm{-1}}$. \emph{J. Phys. Chem. A} \textbf{2008}, \emph{112},
  1480--1492\relax
\mciteBstWouldAddEndPuncttrue
\mciteSetBstMidEndSepPunct{\mcitedefaultmidpunct}
{\mcitedefaultendpunct}{\mcitedefaultseppunct}\relax
\EndOfBibitem
\bibitem[Richardson(2018)]{richardson2018ring}
Richardson,~J.~O. Ring-polymer instanton theory. \emph{Intern. Rev. Phys.
  Chem.} \textbf{2018}, \emph{37}, 171--216\relax
\mciteBstWouldAddEndPuncttrue
\mciteSetBstMidEndSepPunct{\mcitedefaultmidpunct}
{\mcitedefaultendpunct}{\mcitedefaultseppunct}\relax
\EndOfBibitem
\bibitem[Hickmann \latin{et~al.}(2020)Hickmann, Chen, Rotzin, Yang, Urbanski,
  and Avancha]{hickmann:2020}
Hickmann,~B.; Chen,~J.; Rotzin,~M.; Yang,~A.; Urbanski,~M.; Avancha,~S. Intel
  Nervana Neural Network Processor-T (NNP-T) fused floating point many-term dot
  product. 2020 IEEE 27th Symposium on Computer Arithmetic (ARITH). 2020; pp
  133--136\relax
\mciteBstWouldAddEndPuncttrue
\mciteSetBstMidEndSepPunct{\mcitedefaultmidpunct}
{\mcitedefaultendpunct}{\mcitedefaultseppunct}\relax
\EndOfBibitem
\bibitem[K{\"a}ser and Meuwly(2022)K{\"a}ser, and Meuwly]{kaser2022fad}
K{\"a}ser,~S.; Meuwly,~M. Transfer learned potential energy surfaces: accurate
  anharmonic vibrational dynamics and dissociation energies for the formic acid
  monomer and dimer. \emph{Phys. Chem. Chem. Phys.} \textbf{2022}, \emph{24},
  5269--5281\relax
\mciteBstWouldAddEndPuncttrue
\mciteSetBstMidEndSepPunct{\mcitedefaultmidpunct}
{\mcitedefaultendpunct}{\mcitedefaultseppunct}\relax
\EndOfBibitem
\bibitem[Laude \latin{et~al.}(2018)Laude, Calderini, Tew, and
  Richardson]{laude2018ab}
Laude,~G.; Calderini,~D.; Tew,~D.~P.; Richardson,~J.~O. Ab initio instanton
  rate theory made efficient using Gaussian process regression. \emph{Faraday
  Discuss.} \textbf{2018}, \emph{212}, 237--258\relax
\mciteBstWouldAddEndPuncttrue
\mciteSetBstMidEndSepPunct{\mcitedefaultmidpunct}
{\mcitedefaultendpunct}{\mcitedefaultseppunct}\relax
\EndOfBibitem
\bibitem[Fang \latin{et~al.}(2024)Fang, Zhu, Cheng, Hao, and
  Richardson]{newGPR}
Fang,~W.; Zhu,~Y.-C.; Cheng,~Y.-H.; Hao,~Y.-P.; Richardson,~J.~O. Robust
  Gaussian Process Regression method for efficient tunneling pathway
  optimization: Application to surface processes. \emph{J. Chem. Theory
  Comput.} \textbf{2024}, \emph{20}, 3766--3778\relax
\mciteBstWouldAddEndPuncttrue
\mciteSetBstMidEndSepPunct{\mcitedefaultmidpunct}
{\mcitedefaultendpunct}{\mcitedefaultseppunct}\relax
\EndOfBibitem
\bibitem[Redington \latin{et~al.}(2008)Redington, Redington, and
  Sams]{redington2008infrared}
Redington,~R.~L.; Redington,~T.~E.; Sams,~R.~L. Infrared Absorption Spectra in
  the Hydroxyl Stretching Regions of Gaseous Tropolone OHO Isotopomers.
  \emph{Z. Phys. Chem.} \textbf{2008}, \emph{222}, 1197--1211\relax
\mciteBstWouldAddEndPuncttrue
\mciteSetBstMidEndSepPunct{\mcitedefaultmidpunct}
{\mcitedefaultendpunct}{\mcitedefaultseppunct}\relax
\EndOfBibitem
\bibitem[Murdock \latin{et~al.}(2010)Murdock, Burns, and
  Vaccaro]{murdock2010vibrational}
Murdock,~D.; Burns,~L.~A.; Vaccaro,~P.~H. Vibrational specificity of
  proton-transfer dynamics in ground-state tropolone. \emph{Phys. Chem. Chem.
  Phys.} \textbf{2010}, \emph{12}, 8285--8299\relax
\mciteBstWouldAddEndPuncttrue
\mciteSetBstMidEndSepPunct{\mcitedefaultmidpunct}
{\mcitedefaultendpunct}{\mcitedefaultseppunct}\relax
\EndOfBibitem
\bibitem[Li \latin{et~al.}(2019)Li, Evangelisti, Gou, Caminati, and
  Meyer]{caminati:2019}
Li,~W.; Evangelisti,~L.; Gou,~Q.; Caminati,~W.; Meyer,~R. The barrier to proton
  transfer in the dimer of formic acid: a pure rotational study. \emph{Angew.
  Chem. Int. Ed. Engl.} \textbf{2019}, \emph{58}, 859--865\relax
\mciteBstWouldAddEndPuncttrue
\mciteSetBstMidEndSepPunct{\mcitedefaultmidpunct}
{\mcitedefaultendpunct}{\mcitedefaultseppunct}\relax
\EndOfBibitem
\bibitem[Insausti \latin{et~al.}(2022)Insausti, Ma, Yang, Xie, and
  Xu]{insausti2022rotational}
Insausti,~A.; Ma,~J.; Yang,~Q.; Xie,~F.; Xu,~Y. Rotational Spectroscopy of
  2-Furoic Acid and Its Dimer: Conformational Distribution and Double Proton
  Tunneling. \emph{ChemPhysChem} \textbf{2022}, \emph{23}, e202200176\relax
\mciteBstWouldAddEndPuncttrue
\mciteSetBstMidEndSepPunct{\mcitedefaultmidpunct}
{\mcitedefaultendpunct}{\mcitedefaultseppunct}\relax
\EndOfBibitem
\bibitem[Videla \latin{et~al.}(2023)Videla, Foguel, Vaccaro, and
  Batista]{videla2023proton}
Videla,~P.~E.; Foguel,~L.; Vaccaro,~P.~H.; Batista,~V.~S. Proton-Tunneling
  Dynamics along Low-Barrier Hydrogen Bonds: A Full-Dimensional Instanton Study
  of 6-Hydroxy-2-formylfulvene. \emph{J. Phys. Chem. Lett.} \textbf{2023},
  \emph{14}, 6368--6375\relax
\mciteBstWouldAddEndPuncttrue
\mciteSetBstMidEndSepPunct{\mcitedefaultmidpunct}
{\mcitedefaultendpunct}{\mcitedefaultseppunct}\relax
\EndOfBibitem
\bibitem[Bruckhuisen \latin{et~al.}(2021)Bruckhuisen, Dhont, Roucou, Jabri,
  Bayoudh, Tran, Goubet, Martin-Drumel, and
  Cuisset]{bruckhuisen2021intramolecular}
Bruckhuisen,~J.; Dhont,~G.; Roucou,~A.; Jabri,~A.; Bayoudh,~H.; Tran,~T.~T.;
  Goubet,~M.; Martin-Drumel,~M.-A.; Cuisset,~A. Intramolecular H-bond dynamics
  of catechol investigated by THz high-resolution spectroscopy of its
  low-frequency modes. \emph{Molecules} \textbf{2021}, \emph{26}, 3645\relax
\mciteBstWouldAddEndPuncttrue
\mciteSetBstMidEndSepPunct{\mcitedefaultmidpunct}
{\mcitedefaultendpunct}{\mcitedefaultseppunct}\relax
\EndOfBibitem
\bibitem[Tautermann \latin{et~al.}(2003)Tautermann, Voegele, and
  Liedl]{tautermann2003ground}
Tautermann,~C.~S.; Voegele,~A.~F.; Liedl,~K.~R. The ground state tunneling
  splitting of the 2-pyridone{\textperiodcentered} 2-hydroxypyridine dimer.
  \emph{Chem. Phys.} \textbf{2003}, \emph{292}, 47--52\relax
\mciteBstWouldAddEndPuncttrue
\mciteSetBstMidEndSepPunct{\mcitedefaultmidpunct}
{\mcitedefaultendpunct}{\mcitedefaultseppunct}\relax
\EndOfBibitem
\bibitem[Meuwly and Hutson(1999)Meuwly, and Hutson]{MM.morphing:1999}
Meuwly,~M.; Hutson,~J.~M. Morphing ab initio potentials: A systematic study of
  Ne--HF. \emph{J. Chem. Phys.} \textbf{1999}, \emph{110}, 8338--8347\relax
\mciteBstWouldAddEndPuncttrue
\mciteSetBstMidEndSepPunct{\mcitedefaultmidpunct}
{\mcitedefaultendpunct}{\mcitedefaultseppunct}\relax
\EndOfBibitem
\bibitem[Horn \latin{et~al.}(2024)Horn, Vazquez-Salazar, Koch, and
  Meuwly]{MM.morphing:2024}
Horn,~K.~P.; Vazquez-Salazar,~L.~I.; Koch,~C.~P.; Meuwly,~M. Improving
  potential energy surfaces using measured Feshbach resonance states.
  \emph{Sci. Adv.} \textbf{2024}, \emph{10}, eadi6462\relax
\mciteBstWouldAddEndPuncttrue
\mciteSetBstMidEndSepPunct{\mcitedefaultmidpunct}
{\mcitedefaultendpunct}{\mcitedefaultseppunct}\relax
\EndOfBibitem
\bibitem[Sch{\"u}tz and Werner(2001)Sch{\"u}tz, and Werner]{schutz2001low}
Sch{\"u}tz,~M.; Werner,~H.-J. Low-order scaling local electron correlation
  methods. IV. Linear scaling local coupled-cluster (LCCSD). \emph{J. Chem.
  Phys.} \textbf{2001}, \emph{114}, 661--681\relax
\mciteBstWouldAddEndPuncttrue
\mciteSetBstMidEndSepPunct{\mcitedefaultmidpunct}
{\mcitedefaultendpunct}{\mcitedefaultseppunct}\relax
\EndOfBibitem
\bibitem[Sorathia \latin{et~al.}(2024)Sorathia, Frantzov, and
  Tew]{sorathia2024improved}
Sorathia,~K.; Frantzov,~D.; Tew,~D.~P. Improved CPS and CBS extrapolation of
  PNO-CCSD(T) energies: The MOBH35 and ISOL24 data sets. \emph{J. Chem. Theory
  Comput.} \textbf{2024}, \emph{20}, 2740--2750\relax
\mciteBstWouldAddEndPuncttrue
\mciteSetBstMidEndSepPunct{\mcitedefaultmidpunct}
{\mcitedefaultendpunct}{\mcitedefaultseppunct}\relax
\EndOfBibitem
\bibitem[Unke \latin{et~al.}(2024)Unke, St{\"o}hr, Ganscha, Unterthiner,
  Maennel, Kashubin, Ahlin, Gastegger, Medrano~Sandonas, Berryman,
  \latin{et~al.} others]{unke2024biomolecular}
Unke,~O.~T.; St{\"o}hr,~M.; Ganscha,~S.; Unterthiner,~T.; Maennel,~H.;
  Kashubin,~S.; Ahlin,~D.; Gastegger,~M.; Medrano~Sandonas,~L.; Berryman,~J.~T.
  \latin{et~al.}  Biomolecular dynamics with machine-learned quantum-mechanical
  force fields trained on diverse chemical fragments. \emph{Sci. Adv.}
  \textbf{2024}, \emph{10}, eadn4397\relax
\mciteBstWouldAddEndPuncttrue
\mciteSetBstMidEndSepPunct{\mcitedefaultmidpunct}
{\mcitedefaultendpunct}{\mcitedefaultseppunct}\relax
\EndOfBibitem
\bibitem[Unke and Meuwly(2019)Unke, and Meuwly]{MM.physnet:2019}
Unke,~O.~T.; Meuwly,~M. PhysNet: A neural network for predicting energies,
  forces, dipole moments, and partial charges. \emph{J. Chem. Theory Comput.}
  \textbf{2019}, \emph{15}, 3678--3693\relax
\mciteBstWouldAddEndPuncttrue
\mciteSetBstMidEndSepPunct{\mcitedefaultmidpunct}
{\mcitedefaultendpunct}{\mcitedefaultseppunct}\relax
\EndOfBibitem
\bibitem[Mil'nikov and Nakamura(2001)Mil'nikov, and Nakamura]{Milnikov2001}
Mil'nikov,~G.~V.; Nakamura,~H. Practical implementation of the instanton theory
  for the ground-state tunneling splitting. \emph{J.~Chem. Phys.}
  \textbf{2001}, \emph{115}, 6881--6897\relax
\mciteBstWouldAddEndPuncttrue
\mciteSetBstMidEndSepPunct{\mcitedefaultmidpunct}
{\mcitedefaultendpunct}{\mcitedefaultseppunct}\relax
\EndOfBibitem
\bibitem[Frisch \latin{et~al.}(2016)Frisch, Trucks, Schlegel, Scuseria, Robb,
  Cheeseman, Scalmani, Barone, Petersson, Nakatsu~ji, \latin{et~al.}
  others]{gaussian16}
Frisch,~M.; Trucks,~G.; Schlegel,~H.; Scuseria,~G.; Robb,~M.; Cheeseman,~J.;
  Scalmani,~G.; Barone,~V.; Petersson,~G.; Nakatsu~ji,~H. \latin{et~al.}
  Gaussian 16. 2016\relax
\mciteBstWouldAddEndPuncttrue
\mciteSetBstMidEndSepPunct{\mcitedefaultmidpunct}
{\mcitedefaultendpunct}{\mcitedefaultseppunct}\relax
\EndOfBibitem
\bibitem[Smith \latin{et~al.}(2017)Smith, Isayev, and Roitberg]{smith2017ani}
Smith,~J.~S.; Isayev,~O.; Roitberg,~A.~E. ANI-1: an extensible neural network
  potential with DFT accuracy at force field computational cost. \emph{Chem.
  Sci.} \textbf{2017}, \emph{8}, 3192--3203\relax
\mciteBstWouldAddEndPuncttrue
\mciteSetBstMidEndSepPunct{\mcitedefaultmidpunct}
{\mcitedefaultendpunct}{\mcitedefaultseppunct}\relax
\EndOfBibitem
\bibitem[Bannwarth \latin{et~al.}(2019)Bannwarth, Ehlert, and
  Grimme]{bannwarth2019gfn2}
Bannwarth,~C.; Ehlert,~S.; Grimme,~S. GFN2-xTB—An accurate and broadly
  parametrized self-consistent tight-binding quantum chemical method with
  multipole electrostatics and density-dependent dispersion contributions.
  \emph{J. Chem. Theory Comput.} \textbf{2019}, \emph{15}, 1652--1671\relax
\mciteBstWouldAddEndPuncttrue
\mciteSetBstMidEndSepPunct{\mcitedefaultmidpunct}
{\mcitedefaultendpunct}{\mcitedefaultseppunct}\relax
\EndOfBibitem
\bibitem[Larsen \latin{et~al.}(2017)Larsen, Mortensen, Blomqvist, Castelli,
  Christensen, Du{\l}ak, Friis, Groves, Hammer, Hargus, \latin{et~al.}
  others]{larsen2017atomic}
Larsen,~A.~H.; Mortensen,~J.~J.; Blomqvist,~J.; Castelli,~I.~E.;
  Christensen,~R.; Du{\l}ak,~M.; Friis,~J.; Groves,~M.~N.; Hammer,~B.;
  Hargus,~C. \latin{et~al.}  The atomic simulation environment -- a Python
  library for working with atoms. \emph{J. Phys. Condens. Matter}
  \textbf{2017}, \emph{29}, 273002\relax
\mciteBstWouldAddEndPuncttrue
\mciteSetBstMidEndSepPunct{\mcitedefaultmidpunct}
{\mcitedefaultendpunct}{\mcitedefaultseppunct}\relax
\EndOfBibitem
\bibitem[Werner \latin{et~al.}(2019)Werner, Knowles, Knizia, Manby,
  {Sch\"{u}tz}, Celani, Gy\"orffy, Kats, Korona, Lindh, Mitrushenkov, Rauhut,
  Shamasundar, Adler, Amos, Bennie, Bernhardsson, Berning, Cooper, Deegan,
  Dobbyn, Eckert, Goll, Hampel, Hesselmann, Hetzer, Hrenar, Jansen, K\"oppl,
  Lee, Liu, Lloyd, Ma, Mata, May, McNicholas, Meyer, {Miller III}, Mura,
  Nicklass, O'Neill, Palmieri, Peng, Pfl\"uger, Pitzer, Reiher, Shiozaki,
  Stoll, Stone, Tarroni, Thorsteinsson, Wang, and Welborn]{MOLPRO}
Werner,~H.-J.; Knowles,~P.~J.; Knizia,~G.; Manby,~F.~R.; {Sch\"{u}tz},~M.;
  Celani,~P.; Gy\"orffy,~W.; Kats,~D.; Korona,~T.; Lindh,~R. \latin{et~al.}
  MOLPRO, version 2019, a package of ab initio programs. 2019\relax
\mciteBstWouldAddEndPuncttrue
\mciteSetBstMidEndSepPunct{\mcitedefaultmidpunct}
{\mcitedefaultendpunct}{\mcitedefaultseppunct}\relax
\EndOfBibitem
\bibitem[Cs{\'a}nyi \latin{et~al.}(2004)Cs{\'a}nyi, Albaret, Payne, and
  De~Vita]{csanyi2004learn}
Cs{\'a}nyi,~G.; Albaret,~T.; Payne,~M.; De~Vita,~A. “Learn on the fly”: A
  hybrid classical and quantum-mechanical molecular dynamics simulation.
  \emph{Phys. Rev. Lett.} \textbf{2004}, \emph{93}, 175503\relax
\mciteBstWouldAddEndPuncttrue
\mciteSetBstMidEndSepPunct{\mcitedefaultmidpunct}
{\mcitedefaultendpunct}{\mcitedefaultseppunct}\relax
\EndOfBibitem
\end{mcitethebibliography}

\end{document}